\documentclass[conference]{IEEEtran} 
\IEEEoverridecommandlockouts
\usepackage{cite}
\usepackage{amsmath,amssymb,amsfonts}
\usepackage{algorithmic}
\usepackage{graphicx}
\usepackage{textcomp}
\usepackage{xcolor}
\usepackage[acronym,shortcuts]{glossaries}
\usepackage{epstopdf}
\usepackage[caption=false,font=footnotesize]{subfig}
\usepackage{tikz}
\usepackage{caption}
\usepackage[capitalize]{cleveref}
\Crefformat{figure}{#2Fig.~#1#3}
\Crefmultiformat{figure}{Figs.~#2#1#3}{ and~#2#1#3}{, #2#1#3}{ and~#2#1#3}
\usepackage{comment}
\def\BibTeX{{\rm B\kern-.05em{\sc i\kern-.025em b}\kern-.08em    T\kern-.7667em\lower.7ex\hbox{E}\kern-.125emX}}

\def\nfC{\ensuremath{f_\text{C}}}        
\def\nTS{\ensuremath{T_\text{S}}}				 

\begin{document}
\bstctlcite{IEEEexample:BSTcontrol}

\title{Methodologies for Future Vehicular Digital Twins\\
\thanks{\hspace{-1.3\parindent} \textbf{This article is based upon work from COST Action INTERACT, CA20120, supported by COST (European Cooperation in Science and Technology).}\\The work of M. Hofer and T. Zemen is funded within the project DEDICATE (Principal Scientist grant) at the AIT Austrian Institute of Technology. The work of D. Radovic and F. Pasic was supported by the Austrian Research Promotion Agency (FFG) via the research project Intelligent Intersection (ICT of the Future, Grant 880830). The work by A. Fedorov has been realized within the FFI SIVERT 2 project supported by VINNOVA, Sweden's Innovation Agency.}
}

\author{\IEEEauthorblockN{
Danilo Radovic\IEEEauthorrefmark{1},
Markus Hofer\IEEEauthorrefmark{2},
Faruk Pasic\IEEEauthorrefmark{1},
Enrico M. Vitucci\IEEEauthorrefmark{3},
Aleksei Fedorov\IEEEauthorrefmark{6},\\
Thomas Zemen\IEEEauthorrefmark{2},
}%
\IEEEauthorblockA{\IEEEauthorrefmark{1}
Institute of Telecommunications, TU Wien, Vienna, Austria}
\IEEEauthorblockA{\IEEEauthorrefmark{2}
Center for Digital Safety \& Security, AIT Austrian Institute of Technology GmbH, Vienna, Austria}
\IEEEauthorblockA{\IEEEauthorrefmark{3}
Dept. of Electrical and Information Engineering, University of Bologna, Bologna, Italy}
\IEEEauthorblockA{\IEEEauthorrefmark{6}
Dept. of Electrical and Information Technology, Lund University, Lund, Sweden}
\IEEEauthorblockA{	\{danilo.radovic, faruk.pasic\}@tuwien.ac.at, \{markus.hofer, thomas.zemen\}@ait.ac.at, \\ enricomaria.vitucci@unibo.it, aleksei.fedorov@eit.lth.se}
}

\maketitle

\newacronym{AoA}{AoA}{angle of arrival}
\newacronym{AoD}{AoD}{angle of departure}
\newacronym{cir}{CIR}{channel impulse response}
\newacronym{wssus}{WSSUS}{wide-sense stationarity uncorrelated scattering}
\newacronym{lsf}{LSF}{local scattering function}
\newacronym{los}{LOS}{line-of-sight}
\newacronym{nlos}{NLOS}{non-LOS} 
\newacronym{v2i}{V2I}{vehicle-to-infrastructure}
\newacronym{v2v}{V2V}{vehicle-to-vehicle}
\newacronym{v2x}{V2X}{vehicle-to-everything}
\newacronym{mmWave}{mmWave}{millimeter wave}
\newacronym{SDR}{SDR}{software defined radio}
\newacronym{MIMO}{MIMO}{multiple-input multiple-output}
\newacronym{PDP}{PDP}{power delay profile} 
\newacronym{RF}{RF}{radio frequency} 
\newacronym{DSD}{DSD}{Doppler spectral density}
\newacronym{mpc}{MPC}{multipath component}
\newacronym{WSSUS}{WSSUS}{wide-sense stationarity uncorrelated scattering}
\newacronym{RMS}{RMS}{root mean square}
\newacronym{TWDP}{TWDP}{two-wave with diffuse power}
\newacronym{GSCM}{GSCM}{geometry-based stochastic channel model}
\newacronym{DRT}{DRT}{dynamic ray tracing}
\newacronym{RIS}{RIS}{reconfigurable intelligent surface}
\newacronym{TX}{TX}{transmitter}
\newacronym{RX}{RX}{receiver}
\newacronym{SD}{SD}{static discrete}
\newacronym{SNR}{SNR}{signal-to-noise ratio}
\newacronym{ITS}{ITS}{intelligent transportation systems}
\newacronym{FPGA}{FPGA}{field programmable gate array}
\newacronym{ADAS}{ADAS}{advanced driver assistance systems}

\begin{abstract}
The role of wireless communications in various domains of intelligent transportation systems is significant; it is evident that dependable message exchange between nodes (cars, bikes, pedestrians, infrastructure, etc.) has to be guaranteed to fulfill the stringent requirements for future transportation systems. A precise site-specific digital twin is seen as a key enabler for the cost-effective development and validation of future vehicular communication systems. Furthermore, achieving a realistic digital twin for dependable wireless communications requires accurate measurement, modeling, and emulation of wireless communication channels. However, contemporary approaches in these domains are not efficient enough to satisfy the foreseen needs. In this position paper, we overview the current solutions, indicate their limitations, and discuss the most prospective paths for future investigation.   
\end{abstract}


\section{Introduction}
\label{sec:intro}
Safety, dependability, and efficiency are among the key requirements that have to be guaranteed by car manufacturers and traffic engineers.
So far, numerous technical solutions have been applied to fulfill these goals. 
Some of these are electronic stability program (ESP)~\cite{Guo2010}, anti-slip regulation (ASR)~\cite{Guodong2016}, lane assist systems, motion prediction~\cite{Alex2022} and cooperative adaptive cruise control (CACC)~\cite{Milanes2014}. Further performance enhancement can be achieved through expanded situational awareness, which requires information exchange with other traffic participants and the infrastructure, enabled by a dependable wireless communication system~\cite{Milanes2014, Wymeersch2015}.
In terms of vehicular wireless connectivity, either dedicated short-range communications (DSRC) or cellular vehicular to everything (C-V2X) communications are considered as the basis for \ac{ITS}.

However, the growing number of connected users and infrastructure elements (i.e., cars, bikes, pedestrians, traffic signaling infrastructure) leads to new challenges related to increased data volume exchanged wirelessly. 
Specifically, the International Union of Telecommunication (IUT-R) for IMT 2020 sets the requirements of a peak data rate of 20\,Gbit/s, a latency of 1\,ms, and support communication in high-mobility scenarios up to 500 km/h~\cite{3gpp}. 

It is considered that the sub-6\,GHz bands for wireless data transfer could be replaced or complemented by high-frequency ones, as they may offer sufficient bandwidth to enable transmission at high data rates. 
However, the investigation of high-frequency wireless channel characteristics lags behind the traditional sub-6\,GHz channels. 
Moreover, the discussion on future transportation systems has to consider numerous high- or very-high-speed scenarios, which are characterized by the extreme dynamics of the wireless scattering environment, causing rapid changes in the channel statistics over time. 

Evaluation of future \ac{ITS} applications requires comprehensive real-world performance evaluation in vehicular scenarios, characterized by diverse propagation characteristics. However, conducting field measurements in vehicular scenarios is both, cost- and time-intensive, limiting the ability to test performance across a wide range of radio channel conditions. This underscores the need for development of an accurate digital twin, capable of realistically reproducing site-specific channel conditions. Furthermore, a digital twin enables us to overcome one of the major challenges of testing in vehicular scenarios, which is repeatability. In this context, repeatability denotes the ability to test different variations of communication systems over the exact same realistic channel conditions for cross-validation.
Therefore, a digital twin is seen as an enabling methodology of future wireless technologies, offering a virtual representation of site-specific wireless communication link conditions.

The development of a realistic and dependable digital twin is dependent on highly responsive channel emulation and furthermore channel modeling, delivering accurate channel data. The radio propagation modeling is done either using an empirical or deterministic approach. The empirical approach is enabled by accurate channel measurements and subsequent parameter extraction. Further, applications such as virtual drive testing call for a massive, realistic, site-specific data set, which is hardly achievable through channel measurements. Therefore, the channel modeling for these applications is mostly achieved by a deterministic approach and ray tracing in particular. Once again, the accuracy of available ray tracers relies on wireless channel measurements and characterization, which serve as the ground truth to calibrate a ray tracing model. 

All these observations lead us to the identification of four technological challenges (research pillars), which have to be addressed globally on the path to ensure safe and reliable transportation of the future. These research pillars include
\begin{itemize}
    \item the development of advanced channel measurement methods,
    \item efficient parameter extraction and prediction for channel characterization,
    \item effective adaptive modeling of dynamic channels, and
    \item the accurate channel emulation for repeatable and efficient testing, evaluation, and optimization of vehicular communication systems.  
\end{itemize}

In this position paper, we survey the current solutions in these four domains and discuss the most promising research directions in this respect. 

The structure of the paper corresponds to the above-mentioned research pillars.
The paper includes four sections, where these key research directions are described.
\cref{sec:sounding} describes channel measurements and the evolution of measurement frameworks.
Parameter extraction and characterization of vehicular channel measurements are discussed in \cref{sec:extraction}.
\cref{sec:modeling} presents adaptive modeling of highly dynamic channels. Further, \cref{sec:emulation} summarizes the current state of non-stationary, vehicular channel emulation. Current challenges and future directions of technologies that are needed to enable the development of vehicular digital twins are presented in \cref{future:directions}.
Finally,~\cref{sec:conclusion} concludes the paper.

\section{Channel Measurements and Evolution of Measurement Frameworks} \label{sec:sounding}

A comprehensive understanding of vehicular wireless communication channels serves as a fundamental cornerstone in the development of a site-specific digital twin, enabling accurate modeling and simulation of real-world environments.
\Ac{v2x} communication channels are highly dynamic, characterized by rapidly changing channel statistics that vary from scenario to scenario and over time.
 Capturing the real dynamic nature of \ac{v2x} channels requires deploying measurement equipment on moving vehicles that take part in the traffic environment. This introduces logistical complexities, such as battery life and computational resources present in the vehicles, as well as ensuring synchronization among measurement devices. Additionally, the temporal resolution of measurements is constrained by the onsite available data volume. Furthermore, recreating the same or similar channel characteristics for different \ac{v2x} measurement samples over time poses a significant challenge, calling for innovative solutions.

Presently, \ac{v2x} communication mostly considers technologies operating in the sub-6\,GHz frequency domain. The motivation behind this is the technological maturity and cost-effectiveness of these technologies. 
The advance of 5G and 6G, however, brings higher frequency bands in the \ac{mmWave} domain within reach \cite{He2020}. Increased bandwidth required for the exchange of sensor data, as well as the need for sensing capabilities, call for the application of \ac{mmWave} communication technologies, despite their comparatively high path loss and quasi-optic propagation (i.e., little diffraction, small penetration depth) \cite{He2020}. Hence, a comparative study of wireless propagation in sub-6\,GHz and \ac{mmWave} bands based on multi-band measurements is important for model calibration, especially for transmission architectures that utilize diversity across different frequency bands.

\begin{figure}[th!]
\centering
\begin{tikzpicture}
\node[inner sep=0pt] (pdp3) at (0,0)
    {\includegraphics[width=.47\textwidth,trim=0cm 0cm 0cm 0cm, clip]{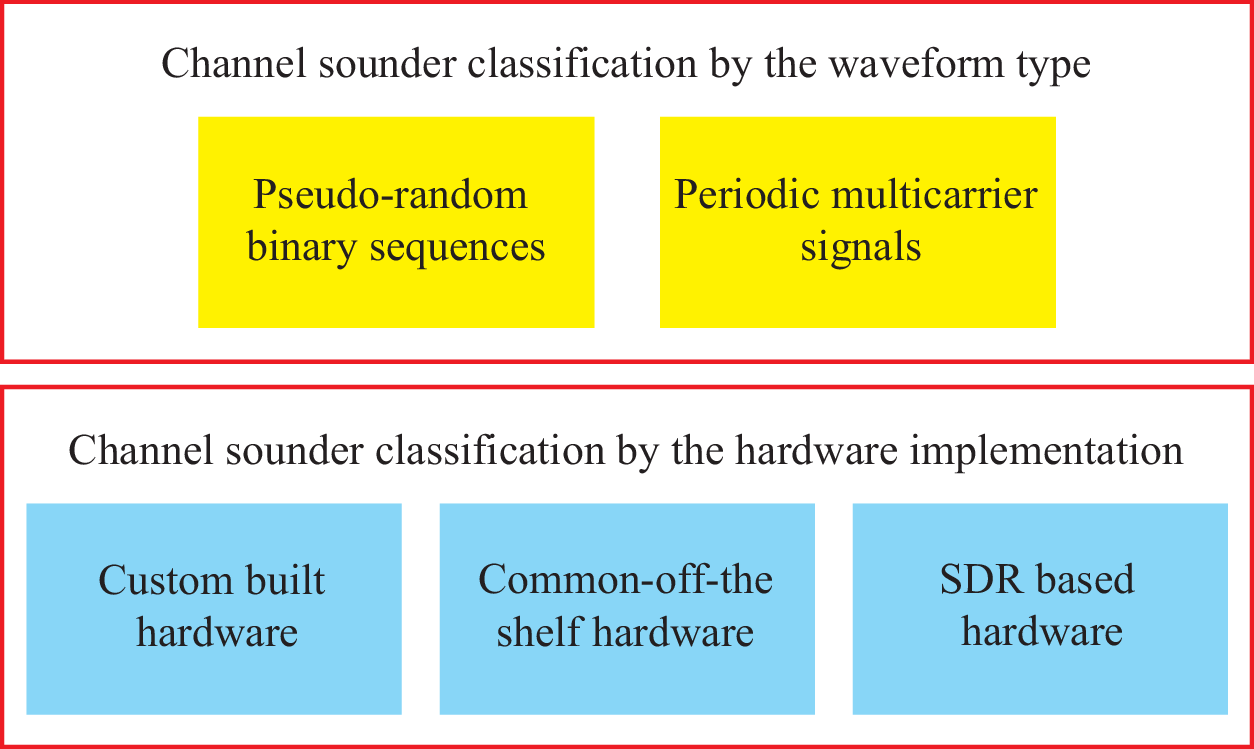}};
    
\end{tikzpicture} 

\caption{Channel sounder classification based on the applied waveform and hardware implementation.} \label{ch_sound_division}
\end{figure}

Currently, correlation-based channel sounders are utilized, where either pseudo-random binary sequences \cite{Park2019,Boban2019,Dupleich2019,Prokes2016} or other types of periodic multicarrier signals \cite{Hofer2021,Hofer2022,Pasic2022,Pasic2021,Pasic2022_2,Radovic2022,Zoechmann2019,Zelenbaba2021,Chopra2022} are used to obtain the \acp{cir} of the wireless communication channels (\Cref{ch_sound_division}). Channel sounders are either custom built \cite{Park2019,Boban2019,Dupleich2019,Chopra2022}, use common-off-the-shelf hardware that is adapted to adhere to the specific needs of the channel measurement campaign \cite{Prokes2016,Pasic2022,Pasic2021,Pasic2022_2,Radovic2022,Zoechmann2019}, or are based on \ac{SDR} \cite{Hofer2021,Hofer2022, Zelenbaba2021} architectures. \ac{SDR}-based channel sounder approaches allow designing highly adaptive channel sounders for multi-node \cite{Zelenbaba2021}, massive \ac{MIMO} or cell-free massive \ac{MIMO} \cite{Loeschenbrand2022}, with affordable costs. Further, to enable the observation of the influence of a single system parameter (frequency band, \ac{SNR}, user velocity, antenna radiation pattern, etc.) while the other parameters remain unchanged, it is essential to develop channel sounders allowing for repeatable measurements.


We introduce two approaches for conducting channel measurements: a) a system with the ability to repeat a measurement under identical channel conditions \cite{Pasic2021} and b) a multi-node channel measurement setup that allows recording the communication channel between multiple nodes simultaneously \cite{Zelenbaba2021} for exactly the same environment.

The channel measurement setup and methodology proposed in \cite{Pasic2021} allow fully automated wireless channel measurements at high speeds and for different frequency bands. The setup is based on a rotary unit, which rotates the omni-directional transmit antenna around a central axis at a constant but adjustable velocity. The rotary unit is equipped with a triggering unit and a rotary joint. The triggering unit generates a trigger pulse at an adjustable arm angle, once per revolution, to initiate the measurement \cite{Pasic2022}. The rotary joint is used to feed the radio frequency signals from a signal source (arbitrary waveform generator) through the rotating arm. The whole rotary unit is placed on a sliding board, that can be moved within the range of 30\,cm along the x-axis and 70\,cm along the y-axis. Thereby, the setup allows for multiple channel realizations within the same environment by conducting measurements at different transmit antenna positions. The testbed setup allows for repeatable and controllable measurements at center frequencies of 2.55, 5.9 and 25.5\,GHz and adjustable velocities of up to 400\,km/h. Therefore, this setup enables performing multiple measurements, where a single observed parameter (frequency band, transmit power, \ac{SNR}, or velocity) is varied while maintaining the remaining aspects of the measurement scenario unchanged. 

The aforementioned rapid fluctuations in channel statistics over time render it impossible to carry out repeatable channel measurements in a real-world street environment. As an alternative solution, it is proposed to measure multiple channels at once, allowing to compare links between different nodes, measured under the same conditions.
The multi-node channel measurement framework presented in \cite{Zelenbaba2021} allows for scalable recording of time-variant frequency responses of multiple wireless links simultaneously for the same measurement run. The fully mobile measurement nodes consist of a \ac{SDR}-based channel sounder with a rubidium clock for synchronization and a power amplifier. Further, this framework has the ability to switch between transmit and receive operation. Results obtained using this framework allow for analysis of delay and Doppler spreads as well as the assessment of available link qualities for multiple communication links. This is of special importance for device-to-device communications systems in vehicular scenarios.

The two presented approaches, the rotating arm within a static environment and the multi-node channel sounder, target the aspect of reproducibility of complex radio measurement problems, e.g., for multi-band or multi-node communication systems. They tackle the challenge of performing repeatable measurements in high-mobility scenarios from two directions. While the first approach is set in laboratory scenarios, allowing for precise control over scenario parameters, the multi-node channel sounder allows for the measurement of the dynamic real-world street environment.

\section{Parameter Extraction and Characterization of Vehicular Channel Measurements} \label{sec:extraction}

For channel modeling and emulation of wireless vehicular channels, multiple parameters such as the size of the channel stationarity region, \ac{RMS} delay spread, \ac{RMS} Doppler spread, $K$-factor and path loss coefficients have to be extracted from measured non-stationary vehicular channels. In \cite{Matz2005} the author introduces a \ac{lsf}. This function is designed for non-stationary channels, as a time-frequency dependent scattering function with a certain size in the time and in the frequency domain, over which the wireless channel is assumed to be approximately \ac{WSSUS}. This time-frequency bounded domain corresponds to a stationarity region. The fulfilled \ac{WSSUS} assumption implies that channel statistics (but not channel realizations) are independent of time and center frequency. The \ac{RMS} delay and Doppler spread are calculated from the \ac{PDP} and \ac{DSD}, which themselves are obtained as the expectation of the \ac{lsf} over the Doppler domain and the delay domain, respectively \cite{Hofer2021, Hofer2022, Bernado2014}. If we denote the \ac{lsf} function, centered at the time and frequency instance $[k_\text{t}, k_\text{f}]$, by $\hat{\mathcal{C}}[k_\text{t}, k_\text{f};l,d]$, we estimate the \ac{PDP} and \ac{DSD} as

\begin{subequations}
    \begin{equation}
    \text{PDP}[k_\text{t},k_\text{f};l] =
   \frac{1}{M}\sum_{d=-M/2}^{M/2-1}\hat{\mathcal{C}}[k_\text{t},k_\text{f};l,d],
    \end{equation}
    \begin{equation}
    \text{DSD}[k_\text{t},k_\text{f};d] =
    \frac{1}{N}\sum_{l=0}^{N-1}\hat{\mathcal{C}}[k_\text{t},k_\text{f};l,d],
    \end{equation}
\end{subequations}
where 
$l$ and $d$ denote \ac{lsf} indices in the delay and Doppler domain, and $N$ and $M$ the size of \ac{lsf} in the delay and Doppler domain, respectively.

The size of the stationarity region depends on the dynamics of the measured scenario, i.e., the movement of transmitter, receiver, and the surrounding environment, as well as the applied center frequency. 
Stationarity investigations using measured sub-6\,GHz vehicular channel impulse response show a strong dependency between stationarity and the observed measurement scenario, based on the estimated stationarity time and bandwidth \cite{ Bernado2012,He2015a}. The authors of \cite{ Bernado2012,He2015a} conclude that the minimum stationarity region is on the order of 100\,ms (corresponding to about 50\,$\lambda$) in the time domain and 40\,MHz in the frequency domain. The work of \cite{Park2019a} evaluates a \ac{v2i} measurement, with the center frequency of 28\,GHz. The measurement is performed in a highway environment, with an approximate speed of 100\,km/h. The results show stationarity times on the order of 2-9\,ms (5-23\,$\lambda$). A first stationarity investigation for measured \ac{v2v} 60\,GHz channels is given in \cite{Radovic2022}. The work demonstrates that the stationarity size in an urban vehicular scenario is on the order of 5-20\,ms (8-32\,$\lambda$) and 270\,MHz in time and frequency, respectively. Furthermore, it is shown that the presence of a dominating \ac{los} signal provides channel stability, leading to larger stationarity region sizes. Dynamic indoor multi-band stationarity investigations demonstrate larger stationarity time in the sub-6\,GHz band compared to the \ac{mmWave} band \cite{Radovic2023}. 
These discussions allow for a first understanding of stationarity in higher frequency bands but also indicate that further repeatable multi-band measurements and evaluations are essential to better understand stationarity region lengths in different vehicular scenarios and between bands.

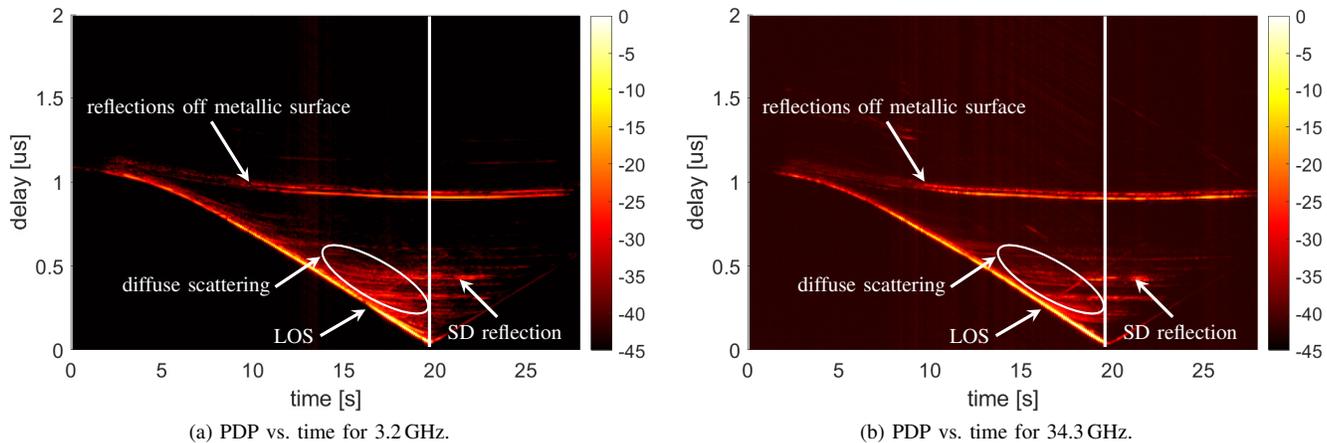
\begin{figure*}[ht!]
\centering
\subfloat[{\ac{PDP} vs. time for 3.2\,GHz.}]{
\begin{tikzpicture}
\node[inner sep=0pt] (pdp3) at (0,0)
    {\includegraphics[width=.47\textwidth,trim=0.5cm 0cm 1.5cm 0cm, clip]{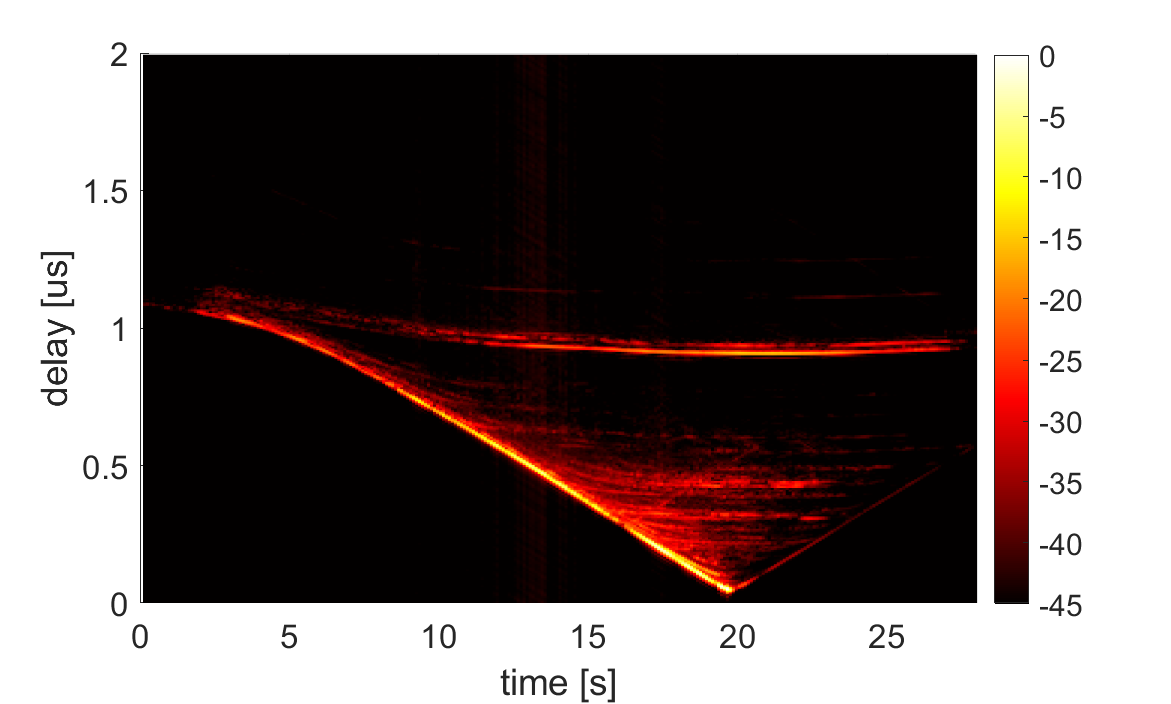}};
    
    \begin{scope}[shift={(0.68,-1.1)},rotate=-30]
    \draw[white, thick]  ellipse (0.8cm and 0.25cm);
    \end{scope}
    \draw [-stealth,very thick, white](-0.65,-1.2) -- (-0.05,-0.8);
    \node at (-1.7,-1.2)[white] {\footnotesize diffuse scattering};
    \draw [-stealth,very thick, white](-0.05,-1.85) -- (0.55,-1.5);
    \node at (-0.4,-1.85)[white] {\footnotesize LOS};
    \node at (-1.4,+1.2)[white] {\footnotesize reflections off metallic surface};
    \draw [-stealth,very thick, white](-1.5,+1) -- (-1,0.2);
    \draw[white, very thick] (1.4,2.4) -- (1.4,-2);
    \draw [-stealth,very thick, white](2.3,-1.6) -- (1.8,-1.1);
    \node at (2.4,-1.8)[white] {\footnotesize SD reflection};
    \label{fig:2a}
\end{tikzpicture}} \quad
\subfloat[{\ac{PDP} vs. time for 34.3\,GHz.}]{
    \begin{tikzpicture}
\node[inner sep=0pt] (pdp34) at (.5\textwidth+0.3cm,0)
    {\includegraphics[width=.47\textwidth,trim=0.5cm 0cm 1.5cm 0cm, clip]{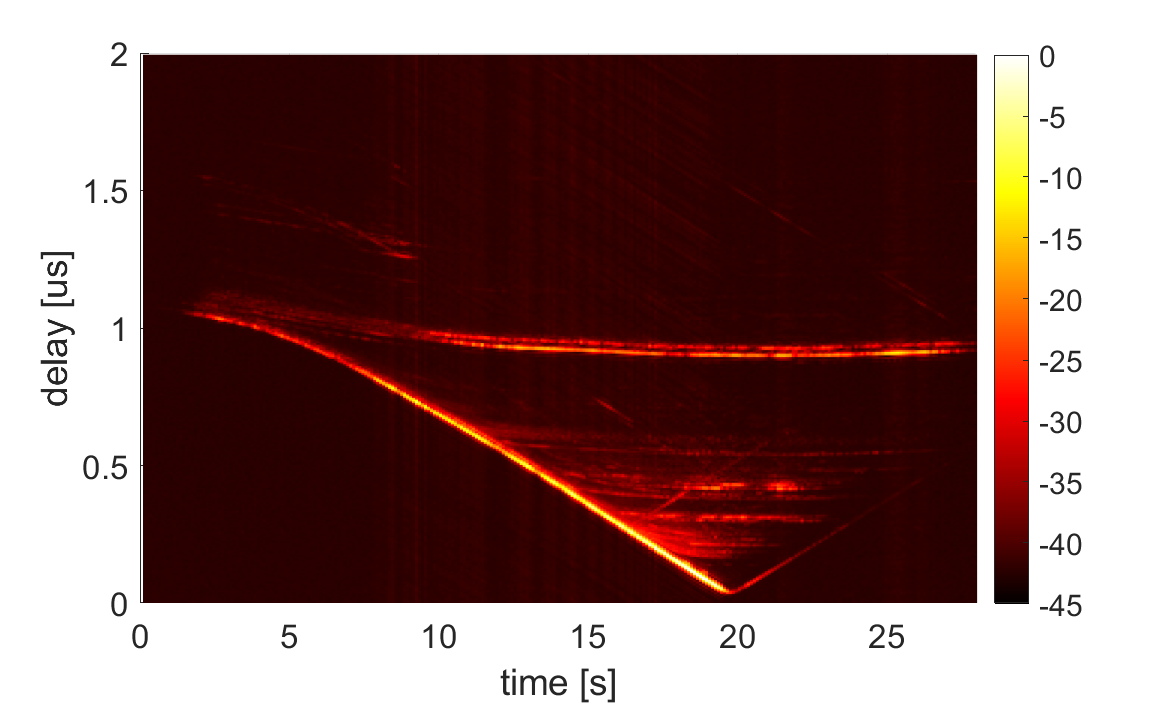}};
 
    \begin{scope}[shift={(10.03,-1.1)},rotate=-30]
    \draw[white, thick]  ellipse (0.8cm and 0.25cm);
    \end{scope}
    \draw [-stealth,very thick, white](8.7,-1.2) -- (9.3,-0.8);
    \node at (7.65,-1.2)[white] {\footnotesize diffuse scattering};
    \draw [-stealth,very thick, white](9.3,-1.85) -- (9.9,-1.5);
    \node at (8.95,-1.85)[white] {\footnotesize LOS};
    \node at (7.95,+1.2)[white] {\footnotesize reflections off metallic surface};
    \draw [-stealth,very thick, white](7.85,+1) -- (8.35,0.2);
    \draw[white, very thick] (10.75 ,2.4) -- (10.75 ,-2);
    \draw [-stealth,very thick, white](11.65,-1.6) -- (11.15,-1.1);
    \node at (11.75,-1.8)[white] {\footnotesize SD reflection};
    
\end{tikzpicture}}

\caption{\ac{v2v} \acp{PDP} of passing car scenario in an urban area for (left) the 3.2\,GHz band and (right) the 34.3\,GHz band. \cite{Hofer2022}}
\label{fig:PDPDSD3p2GHzStreet}
\end{figure*}

The \ac{RMS} delay and the \ac{RMS} Doppler spread in an urban overtaking scenario at the 60\,GHz band are evaluated in \cite{Zoechmann2019}. The authors observe the largest \ac{RMS} delay spread of 4\,ns and argue that it shows resilience to the traffic conditions in the neighboring street lane. Furthermore, the \ac{RMS} Doppler spread is linked to the size of the overtaking vehicles, as the presence of larger trucks notably increases the \ac{RMS} Doppler spread. However, the investigation is influenced by spatial filtering of the used horn antenna and the dominating \ac{los} component. 
\cite{Park2019} provides a delay spread analysis for high-speed \ac{v2i} communication at 28\,GHz. The authors report \ac{RMS} delay spreads of 60\,ns or less, with the \ac{los} component dominating the channel. However, the delay spread is magnified (to 700\,ns or above), when there are \acp{mpc} caused by distant steel structures and surrounding vehicles. These results indicate the dependence of the channel characteristics on the specific scenario and the requirement of multi-band comparison under the same conditions. 
\cite{Dupleich2019} demonstrates strong similarities of multi-band (6.75\,GHz, 30\,GHz and 60\,GHz) delay spread values for synthetic omni-directional measurements. Furthermore, they investigate the dependence of the delay spread on the \ac{TX}-\ac{RX} distance and the distribution of large-scale parameters.

High user mobility, the presence of (static and dynamic) metallic structures in the channel, and low elevated antennas magnify the effect of small-scale fading. Therefore, its characterization for different environment scenarios is of interest. Small-scale fading of the envelope of the \ac{los} tap in the 5\,GHz band may be described by Rician, and later taps by Rayleigh distribution \cite{Bernado2015}. However, in order to obtain accuracy, the $K$-factor has often to be modeled as time and frequency variant. The authors draw the dependence of the $K$-factor on the environment characteristics and variance of the antenna radiation pattern over the bandwidth. For the 60\,GHz urban channels, Rician fading models small-scale fading well in most scenarios \cite{Zoechmann2019}. Nevertheless, the authors argue that in the case when a strong scatterer (e.g. a vehicle) is inside the first Fresnel ellipsoid, small-scale fading is better described by a \ac{TWDP} model. The reasoning is that the \acp{mpc} originating as a reflection of a nearby object arrives at the same delay bin as the LOS component. \cite{wang2018fading} models small scale fading at 73\,GHz as a function of the \ac{TX}-\ac{RX} distance. For the distance under 120\,m small-scale fading follows a Rician distribution. On the other hand, when the \ac{TX}-\ac{RX} distance is in the range of 120-288\,m, small-scale fading is better described by a Nakagami distribution.

To ease the modeling of less investigated \ac{mmWave} channels, it is of significance to investigate the similarities between the frequency bands. Multiple investigations come to a common conclusion that the multi-band measurements are described by very similar \acp{PDP}, suggesting the presence of common scatterers \cite{Dupleich2019,Boban2019,Hofer2021,Hofer2022,Pasic2023}. \cite{Boban2019} and \cite{Dupleich2019} investigate static ultra-wideband multi-band measurements at 6.75\,GHz, 30\,GHz and 60\,GHz demonstrating the vehicle blockage effect of vehicles present in the channel. Further, they show a strong correlation of the \ac{PDP} with the geometry of the channel.

While static measurements offer the advantage of repeatability, to characterize Doppler effects, we have to allow the mobility of \ac{TX}, \ac{RX}, and surrounding vehicles, and measure time variant channels.
 First multi-band sub-6\,GHz - \ac{mmWave} vehicular measurements for dynamic scenarios are given in \cite{Hofer2021} and \cite{Hofer2022}, presenting \ac{v2i} and \ac{v2v} measurements. \Cref{fig:PDPDSD3p2GHzStreet} shows a \ac{PDP} comparison at 3.2\,GHz and 34.3\,GHz for an urban \ac{v2v} scenario, where two vehicles (on which \ac{TX} and \ac{RX} are mounted) drive towards and pass each other in a two-lane street, as depicted in \Cref{fig:AITScen}. The maximum power is normalized to 0\,dB, and the dynamic range is set to 45\,dB. The \ac{los} component dominates the channel in both bands until the \ac{TX} moves out of the \ac{RX} antenna main lobe (around $t=20$\,s). Furthermore, it is to notice the similarity between the measured bands in the distribution of diffuse and \ac{SD} scatterers. 
 In both bands, the channel is strongly influenced by a large metallic surface, present as a facade's metal coating of the far west building in the scenarios, highlighted in yellow rectangular in \Cref{fig:AITScen}.
 
 Considering the aforementioned multi-band comparison, we emphasize the resemblance between the frequency bands. This resemblance encourages the development of cooperative inter-band algorithms.
 Even though analyzing the dynamics of \ac{v2v} channel by observing \ac{PDP} and \ac{DSD} provides an intuitive characterization of the channel resemblance, a detailed analysis of delay spreads is given only for the static scenarios of \cite{Dupleich2019, Boban2019}. Therefore, further exhaustive delay and Doppler spread analyses are necessary to better understand the dynamics of vehicular channels.

\section{Modeling of Highly Dynamic Channels} \label{sec:modeling}

Vehicular communication applications are supposed to demonstrate resilience to high-speed user mobility and consequently rapidly changing scattering environment. Therefore, accurate and realistic modeling of radio communication channels, characterized by rapid fluctuations of channel statistics, is vital for cost-effective development and validation of future vehicular communication systems. 
One of the key considerations in modeling dynamic channels is finding the right trade-off between accuracy and computational complexity.
Optimizing the computation of channel modeling tools can unlock the potential of real-time applications such as human-in-the-loop applications for testing user experience, hardware-in-the-loop applications for testing prototype protocols, and so on.

To reflect the highly dynamic behavior of the channel, a channel simulator must be capable of continuously generating channel realizations, particularly during transitions between different scenarios: moving from a city canyon to an intersection and back, from a highway to a narrow city street, etc. 
Additionally, to accurately represent reality in a vehicular channel model, it should account for the swift variations of the \ac{lsf}.  Currently existing \ac{v2x} simulation frameworks face challenges in modeling the high dynamism of channel evolution, especially when transitioning between different scenarios.  
A couple of challenges are worth mentioning that can be considered as blocking factors for realizing sophisticated channel modeling frameworks: the computational complexity of rich environment modeling, where an enormous amount of objects interact with each other, and the complexity of the conventional ray tracing approach for channel modeling multiplied by the number of objects. 

With the rise in computational resources and enhanced tools for manipulating 3D objects and environments, the V2X community is increasingly drawn to the potential of game engines for channel modeling. Game engines enable developers to implement geometry-based channel models via available "out-of-the-box" ray tracing capabilities. Thus, the Vehicular Networks Simulator with Realistic Physics (VENERIS) channel simulator and its extension~\cite{Veneris, Veneris2} implemented in the Unity game engine offer the state-of-the-art realization of a 3D scene using the OpenStreetMap database (including road infrastructure and buildings) and a GPU-based ray tracing propagation simulator, called Opal. The main drawback of VENERIS is that it assumes complete knowledge about the surrounding's physical and geometrical properties, which is often unavailable except for specific laboratory environments.

\begin{figure*}[ht!]
\centering
\subfloat[{TX and RX pass each other at the indicated passing point.}]{
\begin{tikzpicture}
\node[inner sep=0pt] (UnitySim) at (0,1.0)
    {\includegraphics[width=.47\textwidth]{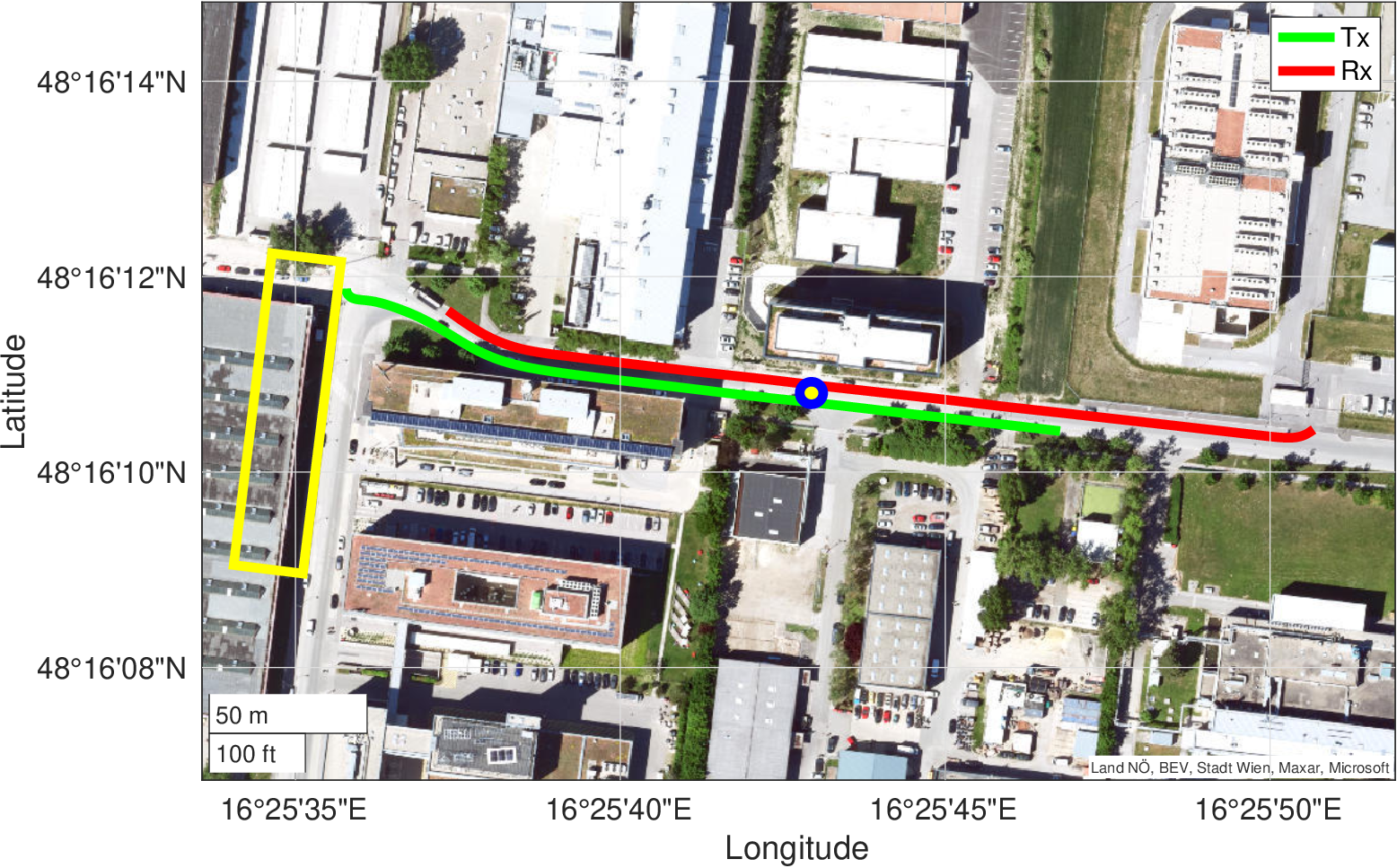}};
    \draw [-stealth,very thick, blue](-1.2,2.5) -- (-2.2,2.0);
    \node at (-1.2,2.6)[draw=black, black,fill=white,very thick] {\footnotesize Metallic surface};
    \draw [-stealth,very thick, blue](1.7,1.8) -- (0.7,1.3);
    \node at (1.7,1.9)[draw=black, black,fill=white,very thick] {\footnotesize Passing point};
    \label{fig:AITScen}
\end{tikzpicture} 
}\quad
\subfloat[{Visualization of a simulation run in Unity near the passing point.}]{
\begin{tikzpicture}
\node[inner sep=0pt] (UnitySim) at (0,1.0)
    {\includegraphics[width=.47\textwidth]{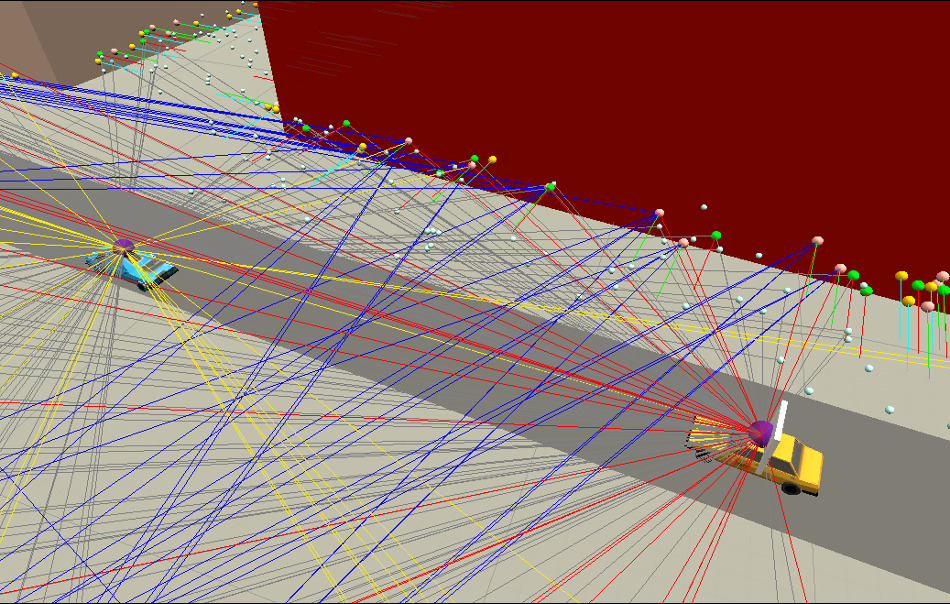}};

    \node at (0.5,+3.0)[white] {\footnotesize $1^\text{st}$ and $2^\text{nd}$ reflection order paths};
    \draw [-stealth,very thick, white](0.5,+2.8) -- (-1,1.2);
    \draw [-stealth,very thick, white](0.5,+2.8) -- (0,1.2);
    \draw [-stealth,very thick, white](0.5,+2.8) -- (1,1.2);
    \draw [-stealth,very thick, white](3.1,-0.8) -- (2.7,-0.3);
    \node at (3.1,-1.0)[white] {\footnotesize \ac{RX}};

    \draw [-stealth,very thick, white](-3.5,2.2) -- (-3.2,1.6);
    \node at (-3.5,2.4)[white] {\footnotesize \ac{TX}};
    \label{fig:Unity3DScenario}
\end{tikzpicture} 
} 
\caption{Importing the real environment (\ref{fig:AITScen}) of the scenario in Fig. \ref{fig:PDPDSD3p2GHzStreet} to a Unity scene (\ref{fig:Unity3DScenario}).}
\label{fig:RealSimSceneTogether}
\end{figure*}
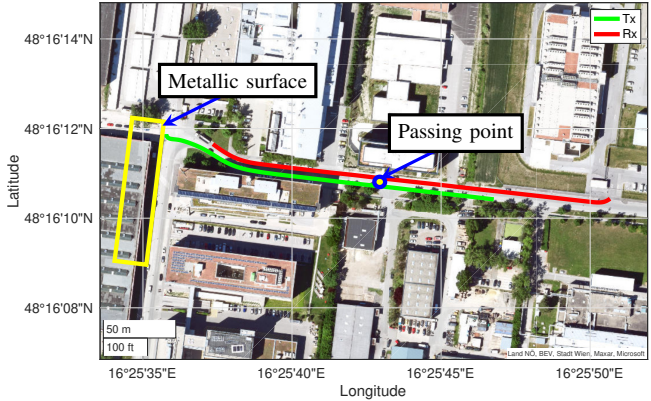
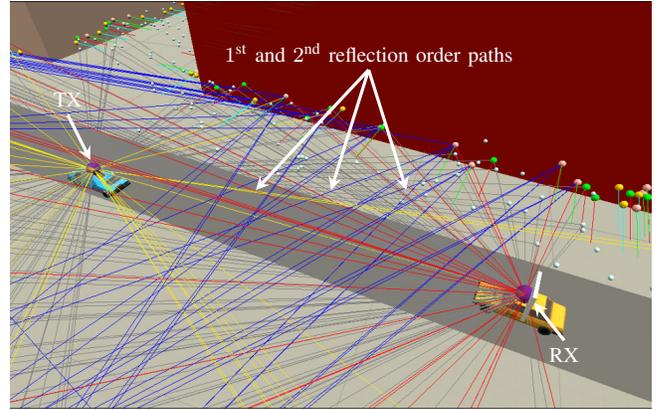

\begin{figure*}[ht!]
\centering
\subfloat[{Simulated \ac{PDP} vs. time for 3.2\,GHz.}]{
    \begin{tikzpicture}
    \node[inner sep=0pt] (pdpSim) at (.5\textwidth+0.3cm,0)
    {\includegraphics[width=.47\textwidth,trim=0.5cm 0cm 1.5cm 0cm, clip]{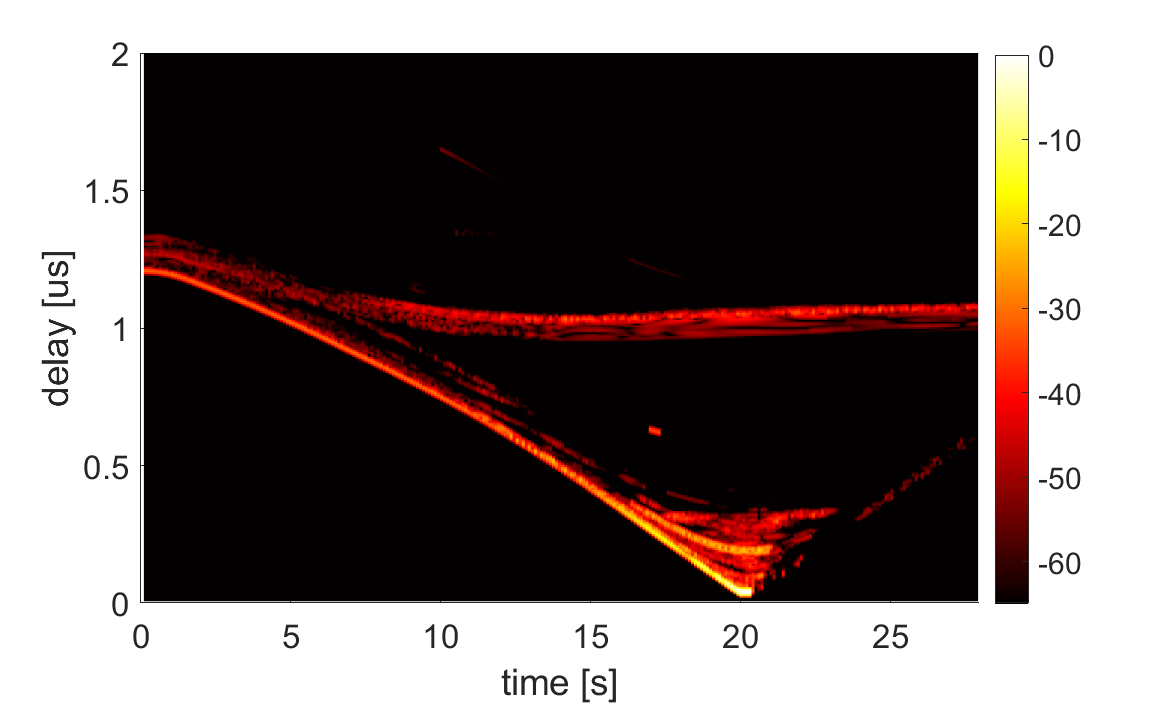}};
 
    \begin{scope}[shift={(10.2,-1.2)},rotate=-30]
    \draw[white, thick]  ellipse (0.8cm and 0.2cm);
    \end{scope}
    \draw [-stealth,very thick, white](8.65,-1.6) -- (9.47,-0.86);
    \node at (8.6,-1.8)[white] {\footnotesize diffuse scattering};
    \draw [-stealth,very thick, white](7.8,-1.0) -- (8.45,-0.45);
    \node at (7.8,-1.2)[white] {\footnotesize LOS};
    \node at (7.85,+1.2)[white] {\footnotesize reflections off metallic surface};
    \draw [-stealth,very thick, white](7.85,+1) -- (8.35,0.3);
    \draw[white, very thick] (10.9,2.4) -- (10.9,-2.1);
    \draw [-stealth,very thick, white](11.9,-0.90) -- (11.45,-1.3);
    \node at (11.77,-0.7)[white] {\footnotesize SD reflection};
\label{fig:ModeledPDP}
\end{tikzpicture}
} \quad
\subfloat[{Simulated vs. measured RMS delay spread for 3.2\,GHz}]{
\begin{tikzpicture}
\node[inner sep=0pt] (UnitySim) at (0,1.0)
    {\includegraphics[width=.47\textwidth]{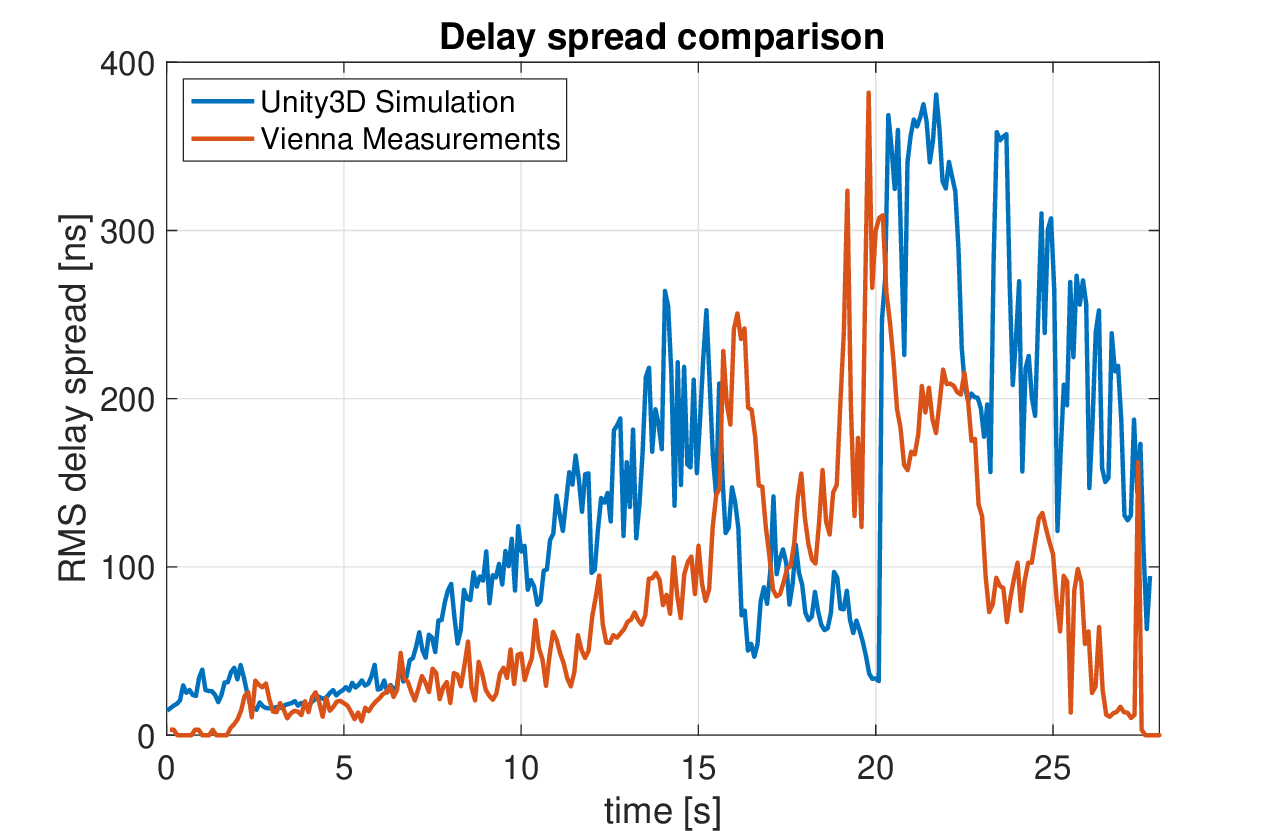}};
    \label{fig:delay_spread}
\end{tikzpicture} 
}
\caption{Example of channel modeling of a scenario with passing cars, matching the scenario in Fig. \ref{fig:PDPDSD3p2GHzStreet}.}
\label{fig:Unity3DSimulation}
\end{figure*}
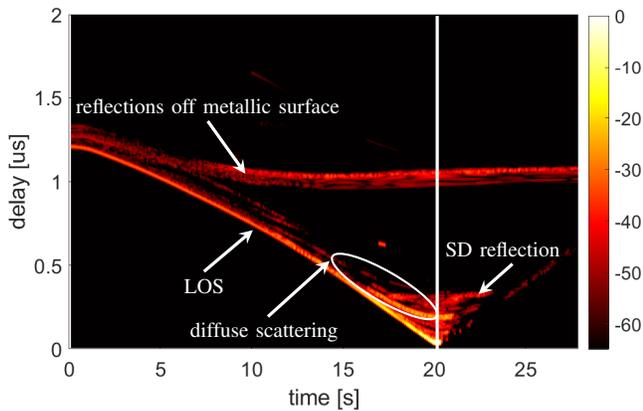
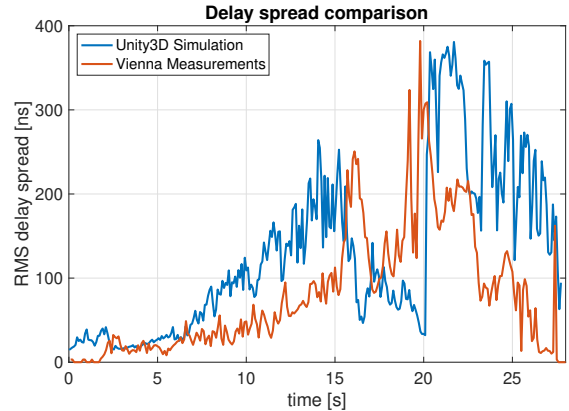


Minor variations in environmental details—such as the opening or closing of windows and doors, the presence of people within the scene, differences in vegetation between summer and winter, and the status of vehicles (whether parked or moving)—can substantially impact signal propagation, thereby affecting the channel realization. Hence, even complete knowledge about the environment should be accompanied by a mechanism that could mimic such variability. To address such kinds of shortages, an approach of \ac{GSCM} has been proposed recently by several academic groups \cite{GSCM, GSCM2, GSCM3, GSCM4}. The idea of a \ac{GSCM} is to outline the surrounding geometry roughly and spread local scatterers around interacting objects to mimic minor variations of the environment based on the geometry-dependent distributions extracted from experimental measurements. The generated channels that consider the combination of geometry and stochastically distributed scatterers can reflect the non-stationary spatio-temporal behavior of wireless channels, both accurately and efficiently~\cite{GSCM, GSCM2, GSCM3}. In other words, the channel's statistics follow vehicles' physical movement through different environments, while keeping spatio-temporal consistency from "snapshot-to-snapshot".

Given the aforementioned advantages, the \ac{GSCM} approach is well-suited for implementation within a game engine. An example of a \ac{GSCM} implementation is illustrated in \Cref{fig:Unity3DScenario} for the Unity game engine. The ray blockage is taken care of by the game engine's built-in functionality of ray tracing, which allows realistic modeling of the scatterers' lifespans. Hence, the variability of the environment governs the LSF, i.e., the appearance and disappearance of local scatterers come naturally from the movement of the vehicles.
A recent \ac{GSCM} implementation in the Unity game engine, similar to the VENERIS framework, has been proposed in \cite{Sivert}. The open-source code is also available in a GitHub repository \cite{SivertUnity}. The main difference of this implementation lies in the ray tracing solution. Instead of launching rays in all directions, as in VENERIS's Opal, rays are sent toward visible scatterers, which decreases the computational complexity of ray tracing. To further optimize the channel model execution, visible scatterers and channel links can be tracked and calculated for all transceiving vehicles in parallel. 
Therefore, achieving real-time execution of channel computation requires an emphasis on key aspects: high parallelization of the channel calculation and effective tracking of visible scatterers. 

\cref{fig:Unity3DSimulation} illustrates a channel calculation example of the scenario with the passing cars (\Cref{fig:AITScen}) in an urban street, matching the scenario description in \cref{fig:PDPDSD3p2GHzStreet}. 
The scenario is implemented using the open-source code \cite{SivertUnity} (see \cref{fig:Unity3DScenario}) without tweaking the original model's parameters. The environment's geometry is extracted from OpenStreetMap.org (further explained in \cite{Sivert, GSCM}). Channel paths of the $1^\text{st}$ and $2^\text{nd}$ reflection order are depicted in gray and yellow/blue/red colors, respectively. For testing purposes, the 2nd reflection order paths are depicted in three colors. The $3^\text{rd}$ reflection order paths are not depicted for the sake of clarity. The resulting \ac{PDP} for the 3.2\,GHz band is presented in \cref{fig:ModeledPDP}. The visual inspection of \acp{PDP} indicates strong similarities between the measured (\Cref{fig:2a}) and modeled (\Cref{fig:ModeledPDP}) channels. 
It is worth emphasizing that channels were generated using the default simulation parameters from the open-source code \cite{SivertUnity}. Consequently, a slight distinction can be observed between the two \acp{PDP}. For clearer visualization of scatterers' contributions in \cref{fig:ModeledPDP}, we adjusted the dynamic range from 45 to 65 dB. This underscores the importance of tailoring GSCM parameters to suit specific scenarios.

As the measurement antennas' radiation characteristics are not known in detail, some other signal propagation effects are also not captured, which can be seen in the segment after the vehicles pass each other. The modeled channel misses catching the \ac{los} component, collected in the back lobe of the \ac{RX} measurement antenna. In the Unity implementation, we emulated the receiving patch antenna (\ac{RX}) from \cite{Hofer2022} as an omni-directional antenna with a white blocking plane behind it, as illustrated in \Cref{fig:Unity3DScenario}, which obviously cannot recreate the same properties of the \ac{RX} antenna. Thus, accounting for antenna radiation patterns is important when modeling the general patterns of channel evolution.

To obtain a more comprehensive comparison between the measured and simulated channel realization, we compare time-varying \ac{RMS} delay spreads (\cref{fig:delay_spread}), calculated by applying the formulas (10) and (11), given in \cite{Bernado2014}. The trend of increased delay spread as the TX and RX come closer together is observable in the measured as well as simulated channels. In the Unity implementation of the \ac{GSCM}, the motion of the vehicles is controlled by the built-in AI Navigation package, which has its own rules on acceleration and collision avoidance. Due to this, vehicles quickly gain high speed initially, coming rapidly to the vicinity of the passing point (\Cref{fig:AITScen}) and swiftly decelerating when they come close to each other to avoid collision. Thus, the \ac{RMS} of the delay spreads significantly differ in the time interval from $t=7$\,s to $t=20$\,s. 
In the \ac{GSCM} implementation, vehicles reach the proximity of the passing point at around 14 seconds and start decelerating, while, in the real measurement campaign, they get the same positions a little later at around 17 seconds. In the time interval from $t=16$\,s to $t=19$\,s, we observe the effect of unadjusted default parameters of the \ac{GSCM} model \cite{SivertUnity}. The reflecting coefficients of the scatterers are obviously smaller than expected. The rapid increase of the delay spread around 20\,s can be explained by the effect of instant disappearance of the \ac{los} component due to blocking plane behind the \ac{RX} antenna.

The provided analyses of the \ac{GSCM} implementation shed light on the drawbacks of the \ac{GSCM} approach. The main drawback lies in the need for the model adjustment for a specific scenario, which can be resource and time-consuming since site-specific measurements must be performed for each new scenario to tune a \ac{GSCM} model. This calls for means of automatic model parametrization, driven by statistics obtained by comprehensive channel measurements and parameter estimation of vehicular scenarios. Also, computational challenges still exist in the \ac{GSCM} approach: the computational complexity increases as $\mathcal{O}(N^L)$, where $N$ refers to the number of scatterers, and $L$ refers to the number of communicating objects (vehicles). This may limit the capability of highly dynamic channel modeling when a huge number of vehicles needs to transit through many scenarios. This emphasizes the requirement for further optimization of the channel modeling.

Another approach meant to mitigate the complexity of modeling channels, characterized by a rapidly changing scattering environment, is given by the emerging paradigm of \ac{DRT} \cite{DRT}.
Since the multipath structure remains essentially constant within a given time-lapse $T_C$, it is possible to predict the multipath evolution on the base of the current multipath geometry. Here, we assume constant speeds and/or accelerations for moving objects within $T_C$, and use analytical extrapolation formulas. In other terms, relying on kinematics theory, it is possible to predict how each reflection/diffraction point "slides" on the interacting wall/edge. This is done without re-running a full ray tracing for every "snapshot" of the environment, therefore with a great computation time-saving, bringing us a step closer to real-time channel modeling. 

\begin{figure}[!th]
    \centering
    \includegraphics[width=0.47\textwidth,trim=0cm 0cm 0cm 3.5cm, clip]{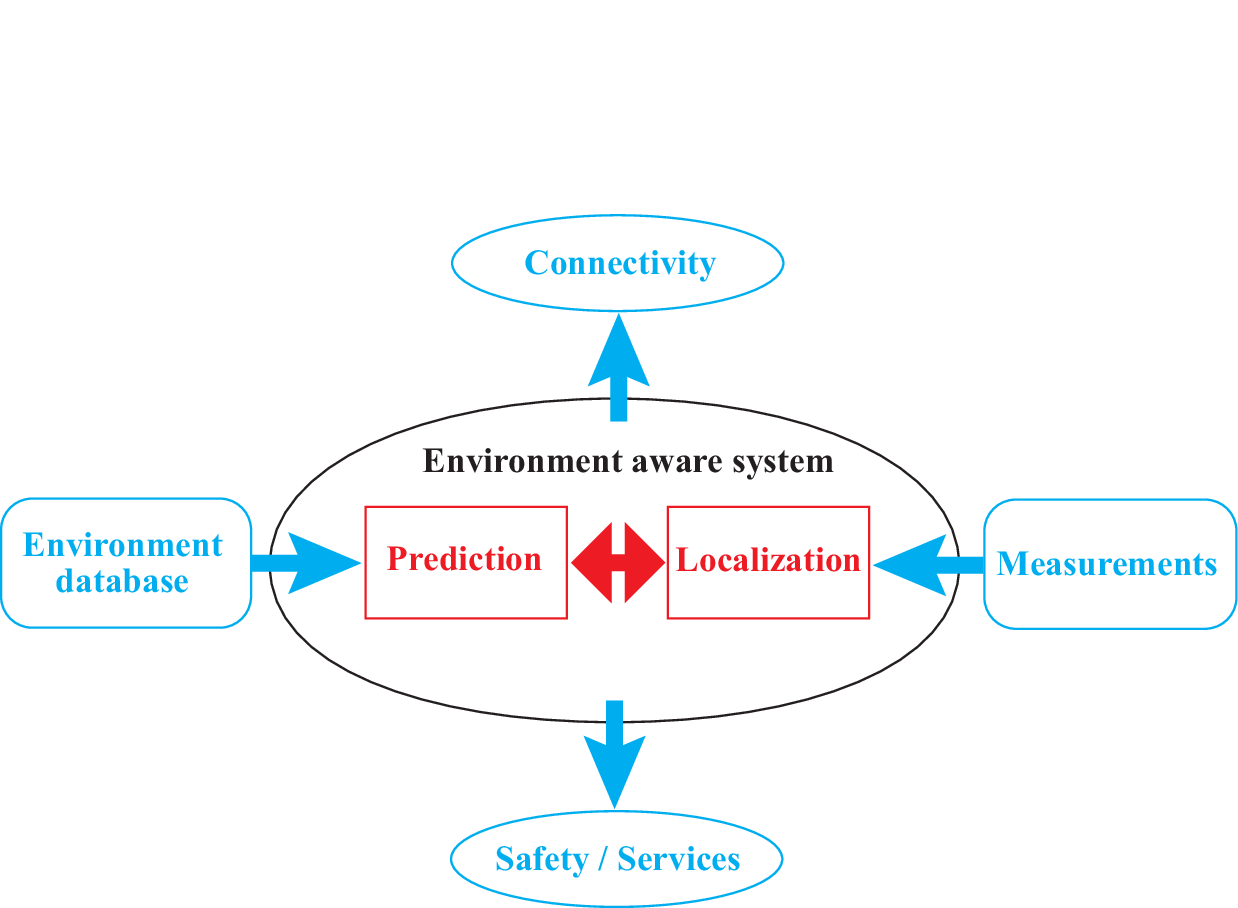}
    \caption{ Scheme of an environment-aware system including both channel prediction and localization.}
    \label{scheme_prediction}
\end{figure}

Further, when \ac{DRT} is embedded in a mobile radio system, ahead-of-time (or anticipative) channel prediction is possible, which opens the way to interesting applications. The anticipative prediction of the environment and/or the radio channel characteristics in high mobility, industrial or vehicular applications allows us to realize the so-called \emph{predictive radio awareness} or \emph{location aware communications}.
Exploiting such capabilities could be of paramount importance to guarantee dependable connectivity in critical applications such as automated and connected driving and to foster interesting safety applications to detect dangerous situations in advance.
Two kinds of applications are possible:
\begin{enumerate}
    \item DRT-based radio channel anticipative prediction
    \item Environment configuration anticipative prediction
\end{enumerate}
In both cases, accurate localization of radio terminals and moving objects, which is likely to be available in future mobile radio systems, is a prerequisite. Accurate localization and the availability of a local environment database are used to build a dynamic environment database for the current time $T_0$.
In 1), \ac{DRT} is used to extrapolate multipath characteristics, and therefore to estimate the channel state information for $t>T_0$. In 2) kinematics theory is used to extrapolate the environment configuration for $t>T_0$ and therefore detect possible hazards or collisions.  These applications will be fully addressed in follow-on studies.
It is worth noting that, in turn, the availability of techniques 1) and 2)  would be of great help to realize multipath-exploiting localization techniques. Therefore, the two goals of anticipative channel prediction and localization might be achieved in synergy to realize an environment-aware system and enhance both connectivity and safety, as depicted in Fig. \ref{scheme_prediction}.

\section{Channel Emulation} \label{sec:emulation}
A digital twin for a radio communication channel relies on a radio channel emulator to carry out hardware-in-the-loop tests (HiL). Therein,  transmit and receive modems are connected to the radio channel emulator to test the communications system repeatedly under well-defined propagation conditions.

The radio channel can be fully described by a sampled time-variant impulse response $h[m,\ell]=h(\nTS m, \nTS \ell)$. Its effect on the transmitted signal is computed by a time-variant convolution
\begin{equation}
y_{r}[m]=\sum_{t=1}^{N_{\text{T}}}\sum_{\ell=0}^{L-1}h_{r,t}[m-\ell,\ell]x_{t}[m-\ell],
\label{eq:convolution}
\end{equation}
where $m$ is the time index, $\ell$ the delay index, $L$ denotes the support of the time-variant impulse response, $r\in\{1,\ldots, N_{\text{R}}\}$ is the receive antenna index, $t$ the transmit antenna index, $N_{\text{T}}$ the number of transmit antennas, $N_{\text{R}}$ the number of receive antennas, and $\nTS $ is the sampling time. The channel impulse response is related to the frequency response via the discrete inverse Fourier transform 
\begin{equation}
h[m,\ell]=1/Q \sum_{q=0}^{Q-1} g[m,q] e^{j 2 \pi \ell q / Q},
\end{equation} 
where $q$ is the frequency index, and $Q$ denotes the number of samples/subcarriers in the frequency domain.

Equation \eqref{eq:convolution} is the fundamental building block of a radio channel emulator and can be implemented by means of digital signal processing. Hence, the radio signal is demodulated and digitized to obtain $x_{t}[m]$. After computing the convolution \eqref{eq:convolution}, the resulting signal $y_{r}[m]$ is digital-analog converted and modulated to the transmit frequency $f_\text{C}$ and fed to the receive modem. So far, channel emulators implement statistical models for the properties of $h[m,\ell]$ using the \ac{PDP}, the \ac{DSD} \cite{Hlawatsch2011} and the spatial correlation for \ac{MIMO} systems. Hence, constant statistical properties are assumed that are \emph{not} fulfilled in vehicular use cases.

The geometry-based radio channel models described in Section \ref{sec:modeling} represent the time-variant channel impulse response as a sum of $P$ propagation paths obtained by the geometric properties of the environment by computing reflection, refraction or diffuse reflection. Linking such channel models to existing hardware channel emulators runs into several limitations. The main limit is the number of delay taps that is limited to $L=48$ for the Keysight Propsim \cite{Mbugua22} and even shorter values for other offers on the market. Groups of 8 taps can be assigned to different regions in the delay domain.

In \cite{Mbugua22} ray tracing data is mapped to the given hardware structure of the Keysight Propsim \cite{PropsimF64} by performing delay alignment. This process avoids the need to interpolate real-valued path delays with the band-limited impulse response of the system. The primary drawback of this approach is the inability to support live updates of the radio channel model while the emulation is running. The live update of the radio channel model requires an additional communication link for the sampled impulse response between the numerical channel model and the convolution unit. The required data rate $R \sim  c_{1} B^2 T_{\text{D}}$ increases quadratically with the bandwidth of the communication system $B=1/\nTS$, where $c_{1}$ is a constant describing the number of bytes per complex sample (e.g. 4), and $T_{\text{D}}$ denotes the support of the delay spread.

A novel channel emulation approach is presented in \cite{Hofer19}. This approach facilitates live updates of the radio channel model while performing channel emulation and thereby achieving a true digital twin. The geometry-based radio channel model structure is used for the radio channel emulation, and the time-variant frequency response is represented as the sum of $P_{s}$ propagation paths
\begin{equation}
g[m,q]= g_{\text{BL}}[q] \sum_{p=1}^{P_{s}}\eta_{p,s} e^{j 2 \pi \nu_{p,s} m'} e^{-j 2 \pi \theta_{p,s} q},
\label{eq:GCM}
\end{equation}
with $m=(M-1)s+m'$, where $M$ is the length of a stationarity region and $s$ denotes the stationarity region index. We drop here the antenna indices $r$ and $t$ for a simpler notation. Within a stationarity region, with a spatial extent $d_{\text{s}}= v \lambda$, where $\lambda=\nfC/c_{0}$ denotes a wavelength, we assume a constant path weight $\eta_{p,s}$ that incorporates the antenna pattern of the \ac{TX} and \ac{RX} antenna, as well as the path loss and attenuation due to reflection on objects. Furthermore, we assume a piecewise constant relative velocity within a stationarity region for path $p$ resulting in a piecewise constant Doppler shift $\nu_{p,s}=f_{p,s}\nTS$, where $f_{p,s}$ is the Doppler shift in Herz. The normalized path delay at $m'=0$ is given by $\theta_{p,s}=\tau_{p}((M-1)s)/{(\nTS Q)}$, where $\tau_{p}(t)$ is the delay of path $p$.

For the real-time implementation, the computation of the complex exponentials in \eqref{eq:GCM} is of high complexity. Hence, \eqref{eq:GCM} is approximated with a reduced rank subspace representation using discrete prolate spheroidal sequences \cite{Slepian78}. It is important to note that the error in terms of MSE can be made arbitrary small by choosing the subspace dimension in time $D_{\text{t}} \ll M$ and frequency $D_{\text{f}} \ll Q$. This is a fundamental advantage compared to the approach in \cite{Mbugua22}. The authors in \cite{Hofer19} use an MSE of $-40\,$dB as design goal. For the projection of each propagation path on this subspace, an efficient algorithm is presented in \cite{Kaltenberger07a} that can be implemented on a \ac{FPGA} by a table lookup and linear scaling. In \cite[Fig. 5.7]{Kaltenberger07} it is shown that the subspace based channel emulation principle is advantageous in terms of complexity for a larger delay spread, diffuse multipath and larger number of propagation paths $P_{\text{s}}$. Here, the complexity of emulating real-time channels is reduced, constituting a crucial step in the process to enable simulations of fast-changing rich scattering environment wireless channels. A careful consideration of the tradeoff between computational complexity and simulation accuracy, based on the available resources and simulation requirements, is to be made.

\section{Future Directions} \label{future:directions}

The development of a site-specific link-level vehicular digital twin relies on advancements in various technological domains. Foremost, further channel measurements and characterization are needed. Within this context, we amplify the need for the classification of characteristic vehicular scenarios and detailed analysis of each one of them, especially in the \ac{mmWave} frequency bands. Additionally, we emphasize the importance of further optimization and computational complexity mitigation in channel modeling and emulation. This optimization is crucial to enable real-time simulations of rich scattering environments, with users moving at high velocity. In this chapter, we go into detail on current challenges and future directions for vehicular digital twin technologies.

\subsection*{Channel Measurements}
Considering the constantly growing demands for higher data throughput and the consequent need for larger bandwidths, future \ac{v2x} channel measurement campaigns should focus on larger bandwidths, particularly within higher frequency bands.
Multi-band channel characterization will be an important aspect, as it is already partially performed within different groups \cite{Boban2019,Dupleich2019,Hofer2021,Hofer2022,Pasic2021,Pasic2022}. The potential relations between bands in terms of power and scatterers can be used, e.g., for beam steering or beam switching. In this aspect, it is crucial to ensure the fair comparability of measurements in terms of dynamic range, antenna pattern, and bandwidth. 

The increase of bandwidth for \ac{SDR} based channel sounders can be obtained using, e.g., frequency stitching methods \cite{Zoechmann2020}. Here, special care has to be taken between the band transitions of neighboring bands to allow for the same power and correct phase relation between the radio bands. Another very important aspect is that much more channel measurement data has to be taken to understand similarities and differences of wireless propagation between bands. This will allow better parameterization, and it will enable automatic parameterization of \acp{GSCM} in the future. 

Future reliable and low-latency \ac{v2i} links in 5G and 6G systems will utilize massive \ac{MIMO} or cell-free massive \ac{MIMO} systems to harvest the channel hardening property of multi-antenna systems to further increase reliability in mobile vehicular scenarios \cite{Loeschenbrand2022}. For these use cases, geometry-based spatially-consistent and frequency-consistent channel modeling methods are required for the research on scalable link-level algorithms and system deployment.

\begin{figure*}[ht!]
\centering
\subfloat[{Highway lane merging scenario.}]{
\begin{tikzpicture}
\node[inner sep=0pt] (pdp3) at (0,1cm)
    {\includegraphics[width=.3\textwidth,trim=0cm 1.3cm 0cm 2cm, clip]{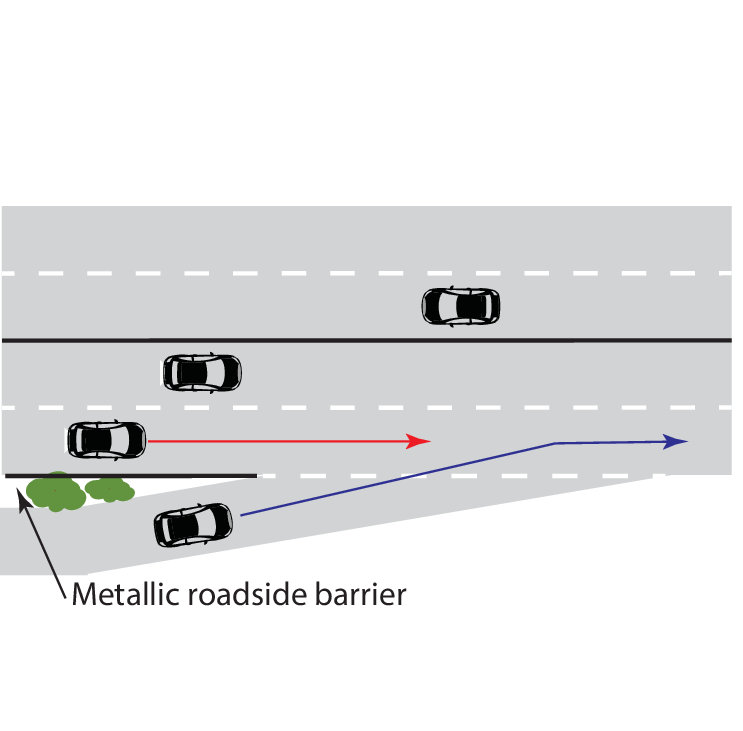}};

\end{tikzpicture} 
} \quad
\subfloat[{Dense traffic vehicle overtaking scenario.}]{
    \begin{tikzpicture}
\node[inner sep=0pt] (pdp34) at (.3\textwidth,1cm)
    {\includegraphics[width=.3\textwidth,trim=0cm -0.5cm 1cm 0cm, clip]{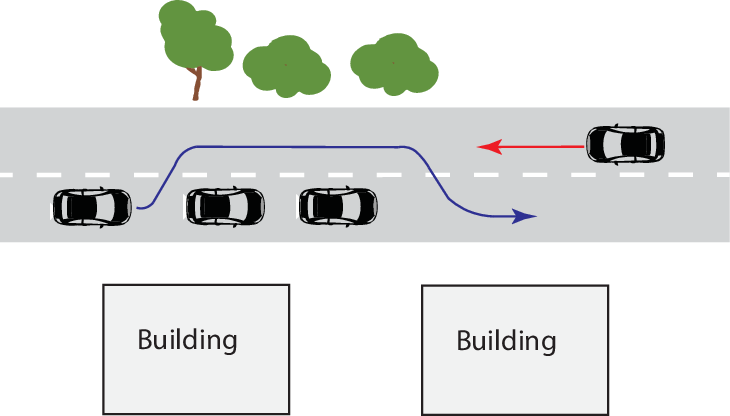}};

\end{tikzpicture}
}\quad
\subfloat[{Urban intersection.}]{
    \begin{tikzpicture}
\node[inner sep=0pt] (pdp34) at (.7\textwidth,0)
    {\includegraphics[width=.3\textwidth,trim=12cm 4.9cm 0cm 1cm, clip]{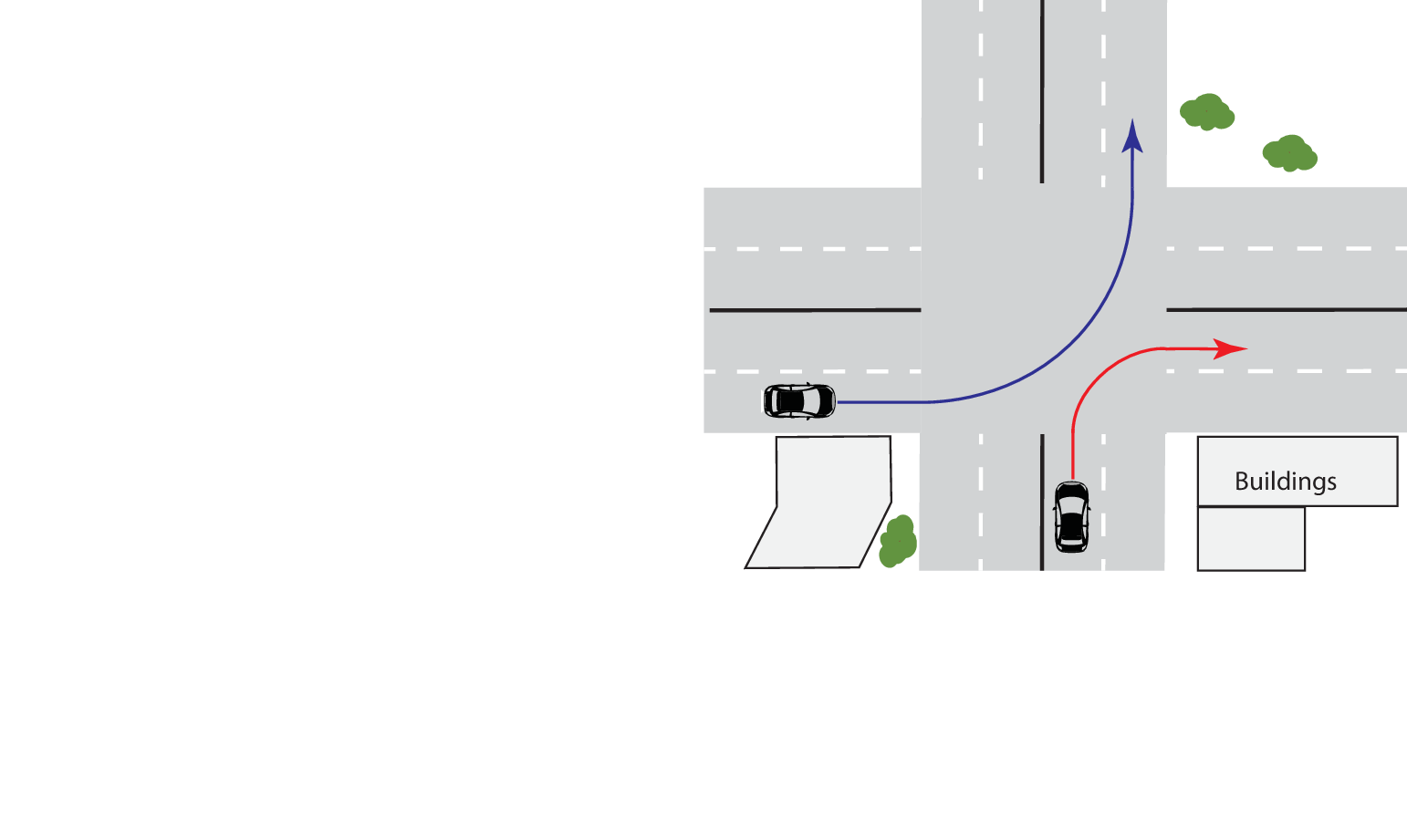}};

\end{tikzpicture}
}
\caption{ Safety-relevant \ac{v2v} communication scenarios. \emph{(vecteezy.com)}}
\label{fig:safety_scenarios}
\end{figure*}

\subsection*{Channel Characterization and Parameter Extraction}
Vehicular communication imposes multiple challenges, which are not present in typical cellular communications. The channel characteristics depend on multiple scenario parameters, e.g., street width, traffic density, speed limit, and street surroundings (building type, vegetation, noise reduction abatement walls, etc.).  Vehicular communication scenarios are typically classified based on the traffic environment on: urban, suburban, highway, rural, etc. \cite{Mecklenbrauker2011}. Due to its peculiarities, the parameters of further specific safety-relevant scenarios (e.g., vehicle overtaking, highway merging lanes, urban intersection) should be characterized in more detail, specifically for multi-band measurements  (\Cref{fig:safety_scenarios}). A first step in this direction for sub-6\,GHz frequency bands has been done in \cite{Bernado2014}. 
Furthermore, low elevated vehicular antennas cause obstruction of a \ac{los} connection in many of the before-mentioned scenarios. Therefore, it is of high interest to investigate the vehicle blockage loss, path loss coefficients, as well as \ac{nlos} communication, relaying on specular \acp{mpc}. 

An additional important aspect is directional channel characterization, specifically when directional antennas or antennas with high antenna gains are utilized to overcome high path loss, as pointed out in \cite{He2020}. 
The \ac{AoA} or \ac{AoD} estimation is usually done by using antenna arrays. However, in vehicular scenarios, we want to avoid placing large, bulky antenna arrays on the vehicles. This can be achieved by using virtual antenna arrays, as proposed in \cite{Zhou2018,Cheng2021}.
However, this approach requires further investigation in the case of non-equidistant measurement points and phase drifts within the measurement sequence. Directional information is important for beam steering, where, e.g., information from lower frequency bands can be used for beam steering in higher frequency bands  \cite{ali2017millimeter}, \cite{burghal2018band} and for channel modeling. Lastly, \acp{RIS} have shown their advantages, improving the performance of vehicular communications \cite{DiRenzo2022,Pan2021}. However, multiple issues (such as the “double fading” effect and  Doppler-induced channel-aging effect) have to be further addressed through channel measurements and characterization.

\subsection*{Channel Modeling}
The main challenge of vehicular channel modeling poses the complexity of numerical modeling fast-changing rich scattering environments.
Furthermore, the modeled channel has to be updated with extreme rates, best described by the requirements for testing communication solutions in the context of \ac{ADAS}. \ac{ADAS} systems are supposed to monitor traffic conditions and swiftly adjust the planned trajectory in response to unexpected road situations \cite{Alieiev2015}. Therefore, the channel modeling has to be responsive enough to capture the wireless channel variations due to these sudden changes in movement trajectories.
In order to address these requirements, optimization of channel modeling is of crucial importance. This optimization can be achieved through the fusion of the current solutions, presented in \cref{sec:modeling}.

The \ac{DRT} method is flexible enough to be employed either in a fully deterministic case or when the path geometry is derived through statistical realizations of the environment, according to the GSCM approach. Therefore, a natural way of future development of highly dynamic channel modeling lies in further optimization of the execution time via merging the \ac{DRT} approach with the game engine-based GSCM framework. In such a way, all the advantages of the previously mentioned approaches - i.e., simplified description of the environment, parallelization on GPUs, and reduction in the number of simulated snapshots - can be exploited and combined together to achieve maximum computational performance and realism in channel modeling of high mobility scenarios.
However, to make this approach more accurate in generating channels and applicable to different scenarios, additional measurements will be needed for different \ac{v2x} environments for further channel characterization. 

Furthermore, the application of the aforementioned modeling methods in the \ac{mmWave} domain is of crucial importance for the development of future vehicular digital twins. However, modeling of \ac{mmWave} channels poses multiple issues, primarily due to the scarcity of available channel characterization. We point out the need for further channel measurements of typical vehicular scenarios, providing a detailed setup description to enable model calibration. Further, as the stationarity time decreases in the \ac{mmWave} domain, real-time modeling requires enhancement of the responsiveness of the model while effectively handling a large number of \acp{mpc}.

\subsection*{Channel Emulation}
Challenging channel emulation scenarios for \ac{v2v} links include urban environments with diffuse multipath and strong, rapid changes leading to a non-stationary fading process, e.g., at street crossings. The channel emulation requirements for the \ac{v2i} link to a base station are influenced by the advancements in 5G advanced and 6G technology, such as:
\begin{itemize}
\item large massive \ac{MIMO} systems for mobile users \cite{Loschenbrand23},
\item cell-free massive \ac{MIMO} for mobile use cases where channel emulation needs to be able to reproduce channel hardening obtained by spatial focussing in cell-free massive \ac{MIMO} \cite{Loschenbrand22}, as well as 
\item joint communication and sensing (JCAS) under multipath conditions.
\end{itemize}
For all three directions, correct geometrical properties of propagation paths are crucial.
Therefore, we emphasize the need for development of further accurate geometry-base radio channel models and matching emulation methods.

\section{Conclusions} \label{sec:conclusion}

In this paper we present a summary of the essential technologies required to facilitate the development of future vehicular link-level digital twins. We examined the current status and future directions of channel measurement, modeling, and emulation of \ac{v2x} communications. The paper showed that the increasing demand for bandwidth has to be considered in the vehicular domain, requiring channel measurement campaigns with higher bandwidths at higher frequency bands. In parameterization, more comparison between the channels in different frequency bands, for specific realistic vehicular scenarios, is of high interest. Furthermore, we discussed the current state and future challenges of dynamic channel modeling. We single out the issue of high computational complexity, required to simulate realistic vehicular channel environments. Therefore, we emphasize further computational optimization of channel modeling and emulation as a key enabler to unlock the potential of real-time applications. 
Moreover, we point out the necessity of obtaining further \ac{v2x} measurements, in order to better understand the underlying statistics. The \emph{Future Direction}, presented in this work, may be used as a guide to shape the vehicular communications of tomorrow, offering valuable insights and recommendations for researchers and developers in the field.

\bibliographystyle{IEEEtran}
\bibliography{lib}

\newpage
\section*{About the Authors}

\begin{IEEEbiography}[{\includegraphics[width=1in,height=1.25in,trim=3cm 10cm 3cm 3cm,clip,keepaspectratio]{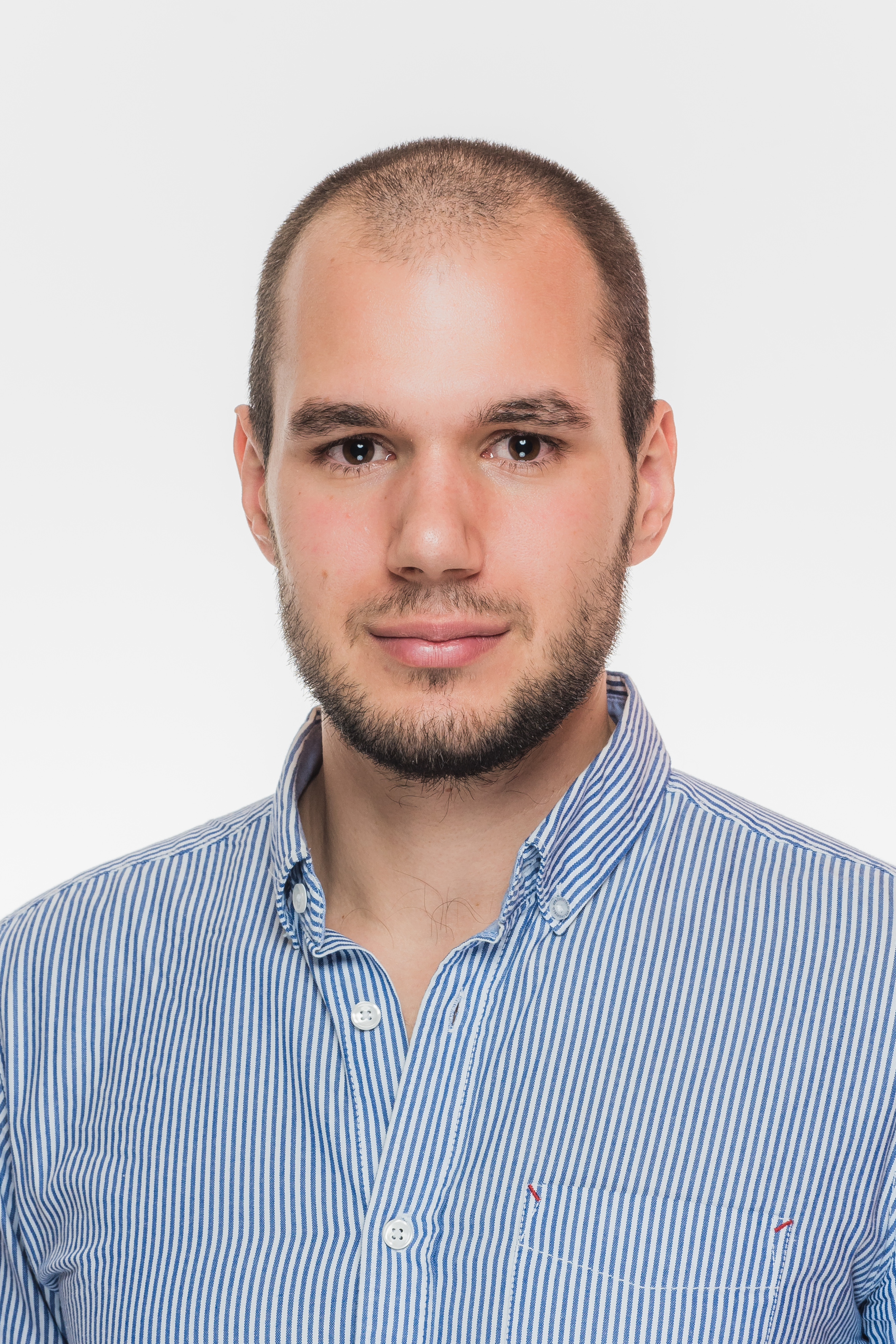}}]{Danilo Radovic}
received the B.Sc. degree in electrical engineering in 2017 and the Dipl.-(Ing.)(M.Sc.) degree in telecommunications in 2019 from TU Wien, Austria. Currently, he is pursuing a Ph.D. degree with the Wireless Communications group at the Institute of Telecommunications, TU Wien. His research focus is on 5G and beyond vehicular wireless communications. The main topics of his research are in the field of wireless channel characterization and vehicular connectivity based on orthogonal time frequency space (OTFS). 
\end{IEEEbiography}

\begin{IEEEbiography}[{\includegraphics[width=1in,height=1.25in,trim=0cm 1.5cm 0cm 0.4cm,clip,keepaspectratio]{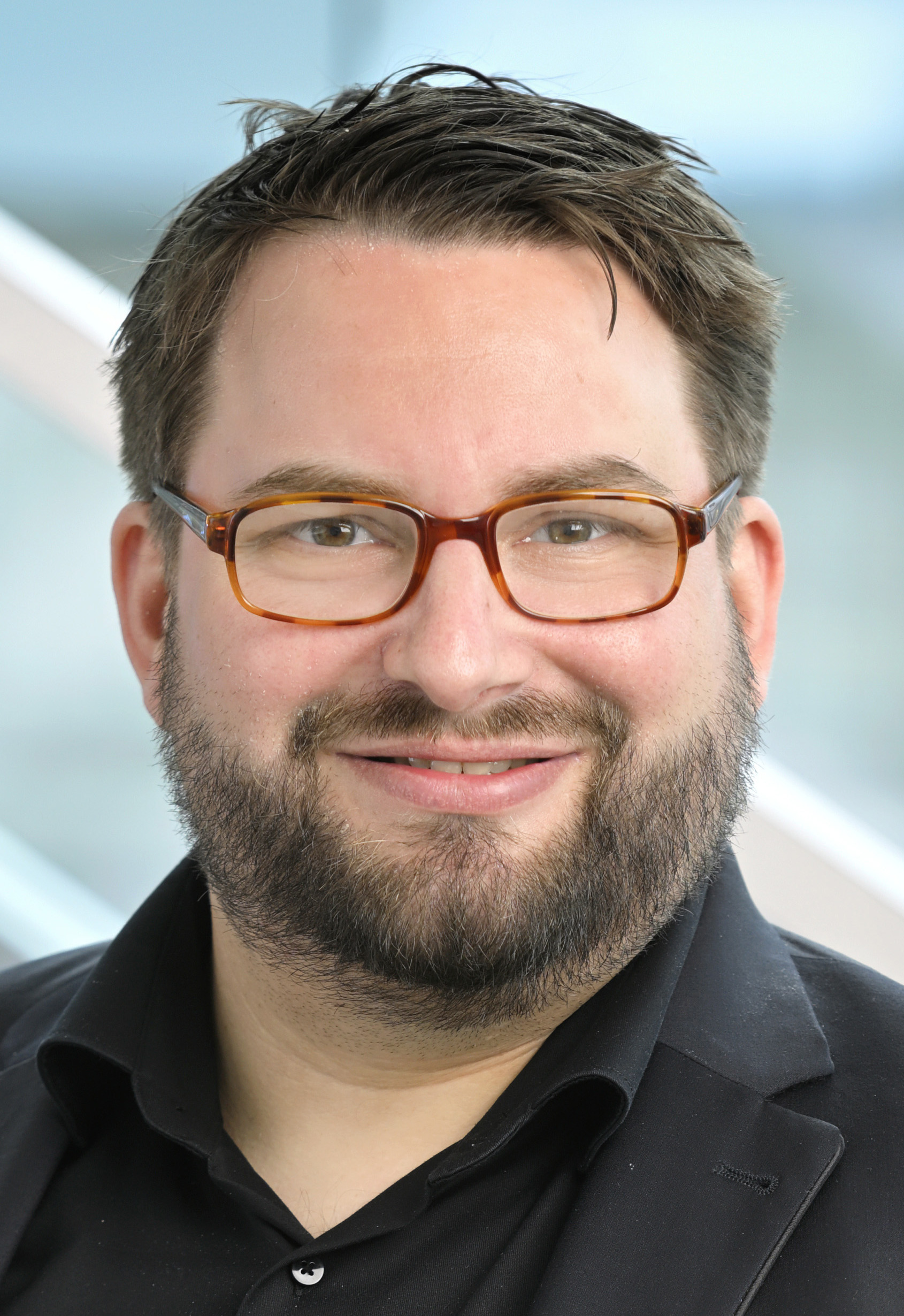}}]{Markus Hofer}
received the Dipl.-(Ing.) degree (Hons.) in telecommunications from the Vienna University of Technology, Vienna, Austria, in 2013, and the Ph.D. degree, in 2019. From 2013 to 2015, he was a Researcher with the Signal and Information Processing Department, FTW Telecommunications Research Center Vienna. He has been with the AIT-Austrian Institute of Technology, Vienna, since 2015. He is currently working as a Scientist with the Research Group for Ultra-reliable Wireless Machine-to-Machine Communications. His research interests include ultra-reliable low latency wireless communications, reﬂective intelligent surfaces, mmWave communications, cell-free massive MIMO, time-variant channel measurements, modeling and real-time emulation, time-variant channel estimation, 5G massive MIMO systems, software-deﬁned radio rapid prototyping, cooperative communication systems, and interference management.
\end{IEEEbiography}

\begin{IEEEbiography}[{\includegraphics[trim=0.5cm 0.0cm 0.5cm 0.1cm,width=1in,height=1.25in,clip,keepaspectratio]{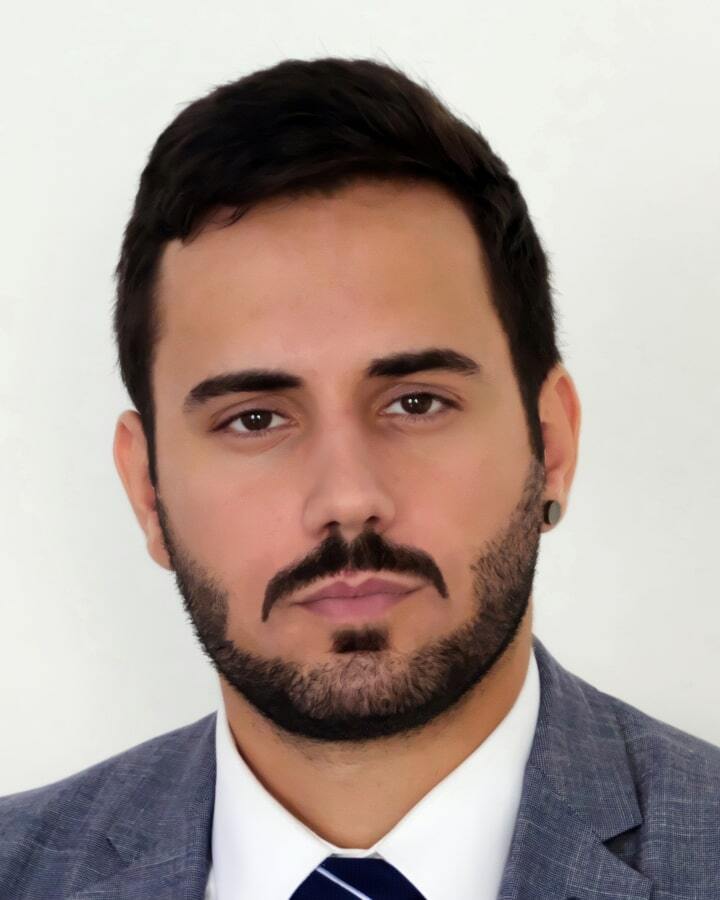}}]{Faruk Pasic} received the B.Sc. degree
in electrical engineering and the Dipl.‑Ing. degree (M.Sc. equivalent) in telecommunications from University of Sarajevo, Bosnia and Herzegovina and TU Wien, Austria, in 2017 and 2021, respectively.
He is currently pursuing the Ph.D. degree in telecommunications engineering with the Institute of Telecommunications, TU Wien.
His main focus is on 5G vehicular-to-everything (V2X) communications.
\end{IEEEbiography}

\vfill

\begin{IEEEbiography}[{\includegraphics[width=1in,height=1.25in,clip,keepaspectratio]{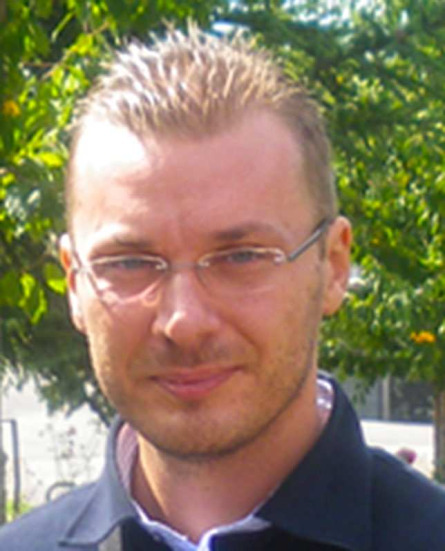}}]{Enrico M. Vitucci} (S’04, M’08, SM’18) is currently an Associate Professor in applied electromagnetics, antennas, and propagation with the Department of Electrical, Electronic and Information Engineering ``G. Marconi'' (DEI), University of Bologna. 
Formerly, he has been a Research Associate with the Center for Industrial Research on ICT, University of Bologna. 
In 2015, he was a Visiting Researcher with Polaris Wireless, Inc., Mountain View, CA, USA.
He is Chair of the Cesena-Forli Unit of the InterDepartment Center for Industrial Research on ICT (CIRI-ICT) of the University of Bologna. He is the author or a coauthor of about 100 technical articles on international journals and conferences, and co-inventor of five international patents.
He participated to several European research and cooperation programs (COST 2100, COST IC1004, COST IRACON, COST INTERACT) and in the European Networks of Excellence NEWCOM and NEWCOM++.
His research interests are in deterministic and wireless propagation models for 5G and beyond. 
Prof. Vitucci is a member of the Editorial Board of the Journal Wireless Communications and Mobile Computing.
\end{IEEEbiography}

\begin{IEEEbiography}[{\includegraphics[width=1in,height=1.25in,clip,keepaspectratio]{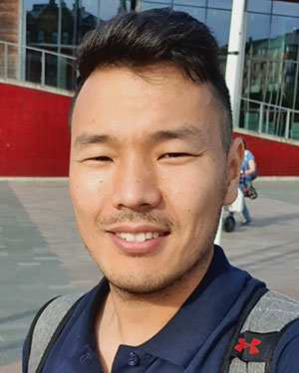}}]{Aleksei Fedorov} is a research engineer specializing
in developing real-time simulators using the Unity game engine for wireless communication systems, including 5G, V2X, and LTE. 
He earned his PhD degree in wireless communications from Otago University in 2019 and holds a Master’s degree in theoretical mechanics from Lomonosov’s Moscow State University  since 2006.
He became a part of the EIT department at Lund University in 2019. 
Currently, his work is centered on creating interactive real-time 3D channel
simulators for V2X, 5G, and emerging 6G applications.
Additionally, Aleksei Fedorov dedicates time to educating the next generation of professionals in the field by lecturing wireless communication-related courses for master’s students at EIT.
His work signifies a balanced mix of academic rigor, practical application, and a commitment to knowledge sharing in the field of wireless communication systems.
\end{IEEEbiography}

\begin{IEEEbiography}[{\includegraphics[width=1in,height=1.25in,trim=0.1cm 0.0cm 0.1cm 0.4cm,clip,keepaspectratio]{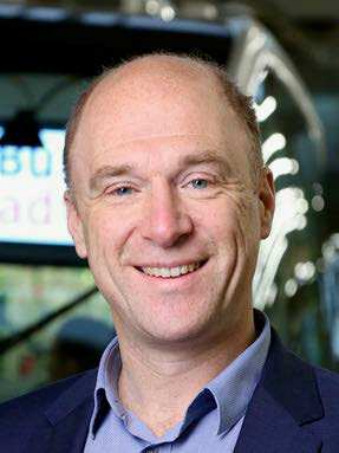}}]{Thomas Zemen}
received the Dipl.-(Ing.) degree (Hons.) in electrical engineering, the Ph.D. degree (Hons.), and the Venia Docendi (Habilitation) degree in mobile communications from the Vienna University of Technology, in 1998, 2004, and 2013, respectively. He joined AIT Austrian Institute of Technology in 2014 and is Principal Scientist since 2021 leading the wireless research group. Previously, Thomas Zemen worked for Siemens AG Austria and the Telecommunication Research Center Vienna (FTW). Mr. Zemen has published four book chapters, 43 journal papers, more than 150 conference communications, and two patents. His research interests are sustainable physical layer radio communication technologies for time-sensitive applications with a focus on 6G technologies for distributed massive MIMO systems, reconfigurable reflective surfaces, multi-band mmWave communications, quantum sensing, and photonic mechanisms for radio frequency systems. Dr. Zemen is docent at the Vienna University of Technology teaching advanced wireless communications.
\end{IEEEbiography}

\vfill

\end{document}